\documentclass[a4paper,10pt,onecolumn,fleqn,oneside]{article}%
\usepackage{graphicx}%
\usepackage{booktabs,rotating,threeparttable}%
\usepackage[sort&compress]{natbib}%
\date{}
\usepackage{amsmath,amssymb}%
\usepackage{subfig}%
\usepackage{textcomp}
\newcommand{\mb}[1]{\mathbf{#1}}%
\ifx\textdegree\undefined%
\newcommand{\textdegree}{\ensuremath{^\circ}}%
\else%
\renewcommand{\textdegree}{\ensuremath{^\circ}}%
\fi%
\begin{document}

\title{On Core and Curvature Corrections used in Straight-Line Vortex Filament Methods\footnote{Submitted for review to the Journal of the American Helicopter Society on November 23, 2011.}}

\author{Wim R. M. Van Hoydonck\\ \small{National Aerospace Laboratory (NLR)}\\\small{1059 CM Amsterdam}\\\small{The Netherlands}\\\small{email: {\tt wim.van.hoydonck@nlr.nl}}\and Marc I. Gerritsma\\ \small{Faculty of Aerospace Engineering}\\\small{Delft University of Technology}\\\small{2629 HT Delft}\\\small{The Netherlands} \and Michel J. L. van Tooren\\ \small{Faculty of Aerospace Engineering}\\\small{Delft University of Technology}\\\small{2629 HS Delft}\\\small{The Netherlands}}

\maketitle

\begin{abstract}%
The convergence characteristics of two viscous core corrections as used in straight-line segmentation methods are rigorously analysed. These are \emph{curvature corrections} that account for the induced velocity contribution at a point on a vortex filament due to local curvature and \emph{core corrections} that remove unrealistically high induced velocities near vortex segments. Two alternative versions of the latter correction are studied: the original as introduced by Scully and a recently introduced improved correction by the authors. The problem is analysed using a viscous vortex ring. A high-order numerical filament method is presented that uses rational B-spline curves to model the vortex ring geometry exactly.
    First, the separate influence of the two corrections on induced velocity values are studied using the numerical filament method. Afterwards, the corrections are combined in four separate cases that are studied using the segmentation method. It is found that in order to get accurate results at coarse discretisations, curvature corrections cannot be neglected. When used together with the original core correction model, results start to diverge from the reference values for discretisations smaller than approximately ten degrees. Combining the curvature correction with the improved core correction extends the discretisation region for which accurate results are found down by approximately one order. Finally, it is shown that segmentation methods that use the original core correction model without curvature correction completely fail to converge to the reference values.
\end{abstract}
\section{Introduction}%
    Engineering rotorcraft simulations that include a free wake component mainly use straight vortex segments to discretise the curved vortices that are trailed from the rotor blades (Refs.~\cite{scully_1975,sadler_1972_1,johnson_camrad_1980}). An analytical solution of the Biot-Savart law is available for this basic wake element (Ref.~\cite{katz_plotkin_book_2001}) which is one reason for its frequent use. A disadvantage of this method is that the analytical solution must be corrected to remove unrealistically high induced velocity values close to the segment.
    
    A consequence of the above approach is that for a marker point on a curved filament, the contributions due to the two directly adjacent segments are zero. In literature, there is disagreement over the need to correct for this contribution at marker points. Bagai (Ref.~\cite{bagai_phd_1995}) and Bhagwat and Leishman (Ref.~\cite{bhagwat_leishman_2001_2}) claim it is not necessary, while others (Refs.~\cite{leonard_1975,scully_1975,saberi_maisel_1987}) claim or implicitely assume it is a contribution that should not be neglected. Other researchers acknowledge that both core corrections and curvature corrections are needed, but then only use the former (Ref.~\cite{egolf_1988}).
    
    The core correction model introduced by Scully (Ref.~\cite{scully_1975}) is used in rotorcraft wake simulation software to remove unrealistically high velocities near a filament. Recently (Ref.~\cite{hoydonck_tooren_2011}), it was shown that when curvature corrections are neglected in induced velocity computations (as in Refs.~\cite{bagai_phd_1995,bhagwat_leishman_2001_2}), the use of Scully's core correction model results in an underprediction of the induced velocity of a viscous vortex ring by approximately 40\%. An improved core correction model was presented that reduces the error considerably at fine discretisations for a large range of relative core sizes. The reference formulas used in Ref.~\cite{hoydonck_tooren_2011} were those for a vortex ring with a non-circular core section (Ref.~\cite{saffman_book_1992}), while in numerical filament methods, a circular core shape is assumed. Recently (Ref.~\cite{hoydonck_gerritsma_tooren_2011}), it has come to our attention that this assumption may lead to different constants in the analytical formulas for the induced velocity of viscous vortex rings (Ref.~\cite{leonard_1985,winckelmans_phd_1989}).
    
    The conclusion in Ref.~\cite{bhagwat_leishman_2001_2} that curvature corrections are superfluous at equivalent arc lengths smaller than approximately 10 degrees is based on the use of the original core correction model. A similar conclusion may not hold when the improved correction is used. 
    
    Therefore, the purpose of this paper is to investigate the convergence characteristics of straight-line segmentation methods when used with 1) the original core correction without curvature correction, 2) the original core correction with curvature correction, 3) the improved core correction without curvature correction and 4) the improved core correction with curvature correction.
    Using this investigation, we would like to give a definite answer to the question whether or not curvature corrections are necessary in straight-line segmentation methods.
\section{Induced Velocity Computations using the Biot-Savart Law}%
    \subsection{Desingularisation of the Biot-Savart Law with Velocity Smoothing Functions}%
        The velocity at a point P induced by a singular vortex filament $\mb{C}(u)$ with strength $\Gamma$ is given by Leonard (Ref.~\cite{leonard_1985}) as
        \begin{equation}\label{eq:biot_savart_param_eqn}
            \begin{split}
                \mb{v}_P &= -\frac{\Gamma}{4\pi} \int_{C(u)} \frac{\mb{r}_P - \mb{r}(u)}{|\mb{r}_P - \mb{r}(u)|^3} \times \frac{\partial \mb{r}(u)}{\partial u} du, \\
                            &= \Gamma \int_{C(u)} \mb{K}(\mb{r}_P - \mb{r}(u)) \times \frac{\partial \mb{r}(u)}{\partial u} du, \\
                            &= \Gamma \int_{C(u)} \Delta \mb{v}(u) du,
            \end{split}
        \end{equation}
        where $\frac{\partial \mb{r}(u)}{\partial u}$ is the derivative at a parametric point $u$ on the filament with respect to the curve parameterisation, $\mb{K}(\mb{r}) = (-1/(4\pi |\mb{r}|^3)) \mb{r}$ is the Biot-Savart kernel, $\mb{r} = \mb{r}_P - \mb{r}(u)$ and $\mb{r}_p$ is the position vector of the point P.
        
        In the vortex dynamics literature (Refs.~\cite{saffman_book_1992,cottet_koumoutsakos_book_2000}), two methods are used to desingularise the Biot-Savart law, the cut-off method and the smoothing method. In the former, a portion of the integral near the point is removed from the integration domain, which removes the singular part. The width of the region that must be excluded is proportional to the relative core size $\sigma/R$. The exact width depends on the vorticity distribution in the core. For example, for a core with a uniform distribution of vorticity, the proportionality constant can be found to be $\frac{1}{2}e^{1/4}$ (Ref.~\cite{saffman_book_1992}, p. 213). This method can only be used for (parametric) points located on the filament and will not be used here. In the smoothing method, the singular Biot-Savart kernel $\mb{K}(\mb{r})$ in Eq.~\ref{eq:biot_savart_param_eqn} is replaced with a mollified kernel $\mb{K}_{\sigma}(\mb{r})$ which removes the singular behaviour at the vortex centerline,
        \begin{equation}\label{eq:mollified_kernel}
            \mb{K}_{\sigma}(\mb{r}) = -\frac{g_{\sigma}(\mb{r})}{|\mb{r}|^3} \mb{r},
        \end{equation}
        where $g_{\sigma}(\mb{r}) = g\left(\frac{|\mb{r}|}{\sigma}\right) = g(\rho)$\footnote{When a smoothing function is displayed without a subscript to denote the dimensionality, it is assumed three-dimensional.} is a three-dimensional velocity smoothing function (Ref.~\cite{winckelmans_leonard_1993}) with $\sigma$ the core size and $\rho$ the relative distance of a point with respect to a point on a filament. The regularised Biot-Savart law can now be written as
        \begin{equation}\label{eq:mollified_biot_savart_param_eqn}
            \mb{v}_P = \Gamma \int_{C(u)} \mb{K}_{\sigma}(\mb{r}_P - \mb{r}(u)) \times \frac{\partial \mb{r}(u)}{\partial u} du = -\Gamma \int_{C(u)} \frac{g(\rho)}{|\mb{r}|^3} \mb{r} \times \frac{\partial \mb{r}(u)}{\partial u} du,
        \end{equation}
        which can be integrated numerically using any suitable quadrature method. The method used to describe the filament geometry is discussed later.
        
        Related to the velocity smoothing is the vorticity smoothing function $\zeta_{\sigma}$ that yields the vorticity field (Ref.~\cite{winckelmans_phd_1989})
        \begin{equation}
            \omega_{\sigma}(\mb{r}) = \Gamma \int_{C(u)} \zeta_{\sigma}(\mb{r}_P - \mb{r}(u)) \frac{\partial \mb{r}(u)}{\partial u} du.
        \end{equation}
        where $\zeta_{\sigma}(\mb{r}) = \frac{1}{\sigma^3}\zeta\left(\frac{|\mb{r}|}{\sigma}\right) = \frac{1}{\sigma^3}\zeta(\rho)$. Given a three-dimensional vorticity smoothing function $\zeta_3(\rho)$, the associated three-dimensional velocity smoothing function $g_3(\rho)$ can be found from (Ref.~\cite{winckelmans_leonard_1993})
        \begin{equation}\label{eq:g3_from_z3}
            g_3(\rho) = \int_0^{\rho} \zeta_3(t) t^2 dt.
        \end{equation}
        
        For an infinitely long, straight filament, the induced velocity field is circumferential. The two-dimensional velocity smoothing $g_2(\rho)$ that induces the same velocity field is obtained from (Ref.~\cite{winckelmans_phd_1989}),
        \begin{equation}\label{eq:g2_from_g3}
            g_2(\rho) = 2\rho^2 \int_{\rho}^{\infty} \frac{g_3(t)}{t^2\sqrt{t^2-\rho^2}} dt.
        \end{equation}
        Similarly, the vorticity distribution in two dimensions $\zeta_2(\rho)$ can be found from (Ref.~\cite{winckelmans_leonard_1993}),
        \begin{equation}\label{eq:z2_from_z3}
            \zeta_2(\rho) = 2 \int_{\rho}^{\infty} \frac{t \zeta_3(t)}{\sqrt{t^2-\rho^2}} dt.
        \end{equation}
        In two dimensions, the velocity and vorticity smoothing functions are related through
        %
        %\begin{equation}
            $g_2(\rho) = \int_0^{\rho} \zeta_2(t) t dt$.
        %\end{equation}
        %
        Lastly, the normalised swirl velocity profile $v_{\theta}(\rho)$ can be found from the two-dimensional velocity smoothing function as follows\footnote{Leishman (Ref.~\cite{leishman_book_2006}) uses $V_{\theta}$, they are related through $V_{\theta} = \frac{\Gamma}{2\pi\sigma}v_{\theta}$.},
        \begin{equation}\label{eq:swirl_velocity_def}
            v_{\theta}(\rho) = \frac{2\pi g_2(\rho)}{\rho}.
        \end{equation}
        It is apparent from Eqs.~\ref{eq:g2_from_g3} and~\ref{eq:z2_from_z3} that computing the two-dimensional smoothings from the three-dimensional smoothings is straightforward, but the opposite can be very difficult.
    \subsection{Example Velocity Smoothing Functions}
        Winckelmans and Leonard (Ref.~\cite{winckelmans_leonard_1993}) give an extensive list of two- and three-dimensional vorticity and velocity smoothing functions with their convergence characteristics. Nowadays, these smoothings are mainly used in applications where a very large set of vortex particles is used to model regions of vorticity. The exact swirl profile induced by a single regularised particle is not as important as the vorticity distribution of a large set of particles. However, when a single parametric curve is used to model a single curved vortex filament (as in helicopter applications), one specific characteristic of the swirl velocity profile is important, namely the location of the extremum. As an example, the swirl velocity profile associated with the high-order algebraic smoothing introduced by Winckelmans and Leonard (Ref.~\cite{winckelmans_leonard_1993}) does not have its extremum at the nondimensional distance $\rho=1$, but more inward (Ref.~\cite{hoydonck_gerritsma_tooren_2011}). For this reason, it is not used here.
        
        Some smoothings relevant to rotorcraft applications are listed in Table~\ref{tab:velocity_smoothings}. These are the Rosenhead-Moore smoothing (Ref.~\cite{leonard_1985}) that yields Scully's swirl velocity profile (Ref.~\cite{scully_1975}), a solid body rotation smoothing (Ref.~\cite{winckelmans_leonard_1993}) that gives the Rankine swirl velocity profile (Ref.~\cite{saffman_book_1992}) and a parametric Gaussian smoothing that yields the Lamb-Oseen swirl velocity profile for $a = 1.2564312$. The velocity smoothings for the recently introduced semi-empirical swirl velocity profile of Ramasamy and Leishman (Ref.~\cite{ramasamy_leishman_2007}) are also given. The three-dimensional velocity smoothing function associated with the Vatistas swirl velocity profile (Ref.~\cite{vatistas_kozel_mih_1991}) for $n=2$ is yet to be found.
    \subsection{Swirl Velocity Smoothing}%
        There is another method in use to construct a smoothing from a swirl velocity profile. Instead of the two-step process $g_3(\rho) \rightarrow g_2(\rho) \rightarrow v_{\theta}(\rho)$ one can skip the middle step and construct a velocity smoothing directly from a swirl velocity profile (Refs.~\cite{bagai_phd_1995} and ~\cite{hoydonck_tooren_2011}). As an example, Scully's swirl velocity profile can be used to construct a three-dimensional velocity smoothing as follows (Ref.~\cite{hoydonck_gerritsma_tooren_2011}),
        \begin{equation}\label{eq:quasi_3d_smoothing_scully}
            v_{\theta}(\rho) = \frac{\rho}{\rho^2+1} \Rightarrow g_3(\rho) = \frac{\rho^2}{4\pi(\rho^2+1)}.
        \end{equation}
        If the equivalent two-dimensional velocity smoothing for the right-hand side of Eq.~\ref{eq:quasi_3d_smoothing_scully} is computed using Eqs.~\ref{eq:g2_from_g3} and~\ref{eq:swirl_velocity_def}, the resulting swirl velocity profile is not the intended one,
        \begin{equation}
            v_{\theta}(\rho) = \frac{\textrm{arcsinh}(\frac{1}{\rho})\rho}{\sqrt{\rho^2+1}}.
        \end{equation}
        A similar exercise can be done for other velocity profiles, but not all of them result in closed-form expressions for the final swirl velocity profile. For example, when constructing a three-dimensional smoothing from the Vatistas $(n=2)$ swirl velocity profile,
        \begin{equation}\label{eq:quasi_3d_smoothing_vatistas}
            v_{\theta}(\rho) = \frac{\rho}{\sqrt{\rho^4+1}} \Rightarrow g_3(\rho) = \frac{\rho^2}{4\pi\sqrt{\rho^4+1}},
        \end{equation}
        the resulting swirl profile is expressed in terms of the generalised hypergeometric function $_3F_2$ (Ref.~\cite{dlmf_nist}),
        \begin{equation}\label{eq:quasi_3d_smoothing_vatistas_swirl_vel}
            v_{\theta}(\rho) = \frac{_3F_2(\frac{1}{2},\frac{1}{2},1;\frac{3}{4},\frac{5}{4};\frac{-1}{\rho^4})}{\rho},
        \end{equation}
        which has an extremum of approximately 0.8845 at the non-dimensional distance $\rho=0.7085$.
        
        This method of directly constructing a smoothing from a swirl velocity profile is not used in the rest of this paper as it results in incorrect swirl velocity profiles.
    \subsection{Velocity of a Viscous Vortex Ring}\label{subsec:viscous_ring_velocity}
        When the regularised Biot-Savart law (Eq.~\ref{eq:mollified_biot_savart_param_eqn}) is used to compute the induced velocity at a point located at the center of the core of a viscous vortex ring, the following asymptotic formula in terms of the three-dimensional velocity smoothing $g_3(\rho)$ can be derived (Ref.~\cite{leonard_1985}),
        \begin{equation}\label{eq:vortex_ring_3d_smoothing}
            U_{g_{3}} = \frac{\Gamma}{4\pi R}\left[\log\left(\frac{4R}{\sigma}\right) - 4\pi\int_0^{\infty} g_3'(\rho)\log \rho d\rho\right],
        \end{equation}
        with $g_3'(\rho) = \rho^2 \zeta_3(\rho)$. However, this formula is derived under the assumption that the shape of the core is circular. More careful analysis shows that the shape of the core is a slightly deformed circle which ultimately leads to (Refs.~\cite{fraenkel_1970,saffman_1970})
        \begin{equation}\label{eq:vortex_ring_2d_smoothing}
            U_{g_2} = \frac{\Gamma}{4\pi R} \left[\log\left(\frac{8R}{\sigma}\right) - \frac{1}{2} + \left(\int_0^1(2\pi g_2(\rho))^2 \frac{d\rho}{\rho} + \int_1^{\infty} \left((2\pi g_2(\rho))^2-1 \right)\frac{d\rho}{\rho} \right)\right].
        \end{equation}
        \subsubsection{Examples}
        When Eqs.~\ref{eq:vortex_ring_3d_smoothing} and~\ref{eq:vortex_ring_2d_smoothing} are used with the corresponding three- and two-dimensional smoothings, the results are in general not the same, except for the Rosenhead-Moore kernel. Substituting its velocity smoothing in Eq.~\ref{eq:vortex_ring_3d_smoothing} leads to 
        \begin{equation}\label{eq:vr_rosenhead_moore}
            U_{g_3}^{RM} = \frac{\Gamma}{4\pi R} \left[ \log \left( \frac{8R}{\sigma}\right) - 1\right],
        \end{equation}
        where $R$ is the radius of the vortex ring and $\sigma$ the vortex core radius. The same result is found when the two-dimensional velocity smoothing is substituted in Eq.~\ref{eq:vortex_ring_2d_smoothing}.
        
        For the solid body rotation kernel, the results do not agree. Substituting the three-dimensional smoothing into Eq.~\ref{eq:vortex_ring_3d_smoothing} yields
        \begin{equation}\label{eq:vr_numerical_rankine}
            U_{g_3}^{SB} = \frac{\Gamma}{4\pi R} \left[ \log \left( \frac{8R}{\sigma}\right) - \frac{1}{2}\right].
        \end{equation}
        Only when the two-dimensional velocity smoothing is used, one gets the correct formula as first derived by Kelvin (see e.g. Saffman, Ref.~\cite{saffman_book_1992}),
        \begin{equation}\label{eq:vr_theoretical_rankine}
            U_{g_2}^{SB} = \frac{\Gamma}{4\pi R} \left[ \log \left( \frac{8R}{\sigma}\right) - \frac{1}{4} \right].
        \end{equation}

        A vortex ring with the two-dimensional parametric Gaussian smoothing has a velocity of
        \begin{equation}\label{eq:vr_theoretical_lamb_oseen}
            U_{g_2}^{PG} = \frac{\Gamma}{4\pi R} \left[ \log\left(\frac{8R}{\sigma}\right) - \left( \frac{1}{2} - \frac{\gamma}{2} + \frac{1}{2}\log\frac{2}{a} \right) \right],
        \end{equation}
        where $\gamma$ is the Euler-Mascheroni constant (Ref.~\cite{abramowitz_stegun_book_1970}). When the equivalent three-dimensional smoothing is used, the result is slightly different,
        \begin{equation}\label{eq:vr_numerical_lamb_oseen}
            U_{g_3}^{PG} = \frac{\Gamma}{4\pi R} \left[\textrm{log}\left(\frac{8R}{\sigma}\right) - \left(1 - \frac{\gamma}{2} + \frac{1}{2}\log\frac{1}{a} \right)\right].
        \end{equation}

        Under the assumption that the core has a circular shape with vorticity distribution as given by the Ramasamy-Leishman model, 
        Eq.~\ref{eq:vortex_ring_3d_smoothing} can be found to give
        \begin{equation}\label{eq:vr_numerical_ramasamy_leishman}
            U_{g_3}^{RL} = \frac{\Gamma}{4\pi R} \left[ \log\left(\frac{4R}{\sigma}\right) - \sum_{n=1}^{3} a_n\left(1-\frac{\gamma}{2} - \frac{1}{2}\log(4 b_n)\right) \right].
        \end{equation}
        Values for the constants $a_n$ and $b_n$ ($n \in 1,\ldots,3$) are listed in Ref.~\cite{ramasamy_leishman_2007}.
        The formula for the correct induced velocity (using Eq.~\ref{eq:vortex_ring_2d_smoothing}) contains divergent terms for values of $a_n \ne 1$. This means that it is only available for low Reynolds numbers, and given by Eq.~\ref{eq:vr_theoretical_lamb_oseen} where $a$ is replaced with the appropriate value of $b_1$.
        
        Since the three-dimensional velocity smoothing associated with the Vatistas swirl velocity profile is not known, no solution is available for Eq.~\ref{eq:vortex_ring_3d_smoothing}. However, the theoretically correct velocity (Eq.~\ref{eq:vortex_ring_2d_smoothing}) for a vortex ring with the Vatistas ($n=2$) swirl velocity is known,
        \begin{equation}\label{eq:vr_theoretical_vat2}
            U_{g_2}^{V2} = \frac{\Gamma}{4\pi R} \left[\log\left(\frac{8R}{\sigma}\right) - \frac{1}{2} \right],
        \end{equation}
        which is the same as the result obtained with the three-dimensional solid-body rotation smoothing (Eq.~\ref{eq:vr_numerical_rankine}).
        
        In this investigation, it is assumed that the filament core is circular in shape. When validating numerically computed induced velocity values, this assumption is also used for the reference formulas. Therefore, the formulas derived in this section that assume a circular core geometry will be used as reference instead of the correct formulas that use no such assumption. The error that is made by this assumption can be quantified by computing the difference between the resulting formulas for a certain core model.  
         
        For a medium-size transport helicopter in hover, typical values of $\frac{\Gamma}{4\pi R}$ are $O(0.1)$, given that the strength of the tip vortices can be computed with $\Gamma_{hov} = \frac{C_T 2\pi R^2 \Omega}{n_{bl}}$ (Ref.~\cite{meijer_drees_1949}).
        For the Rankine core, the difference between Eq.~\ref{eq:vr_numerical_rankine} and Eq.~\ref{eq:vr_theoretical_rankine} is then on the order of 0.025 m/s. A similar difference can be found when the two equations for the velocity of a vortex ring with a laminar core are compared (Eqs.~\ref{eq:vr_theoretical_lamb_oseen} and~\ref{eq:vr_numerical_lamb_oseen}). This small difference shows that the use of vortex filaments with circular cores in rotorcraft wake simulations does not give a large error given that absolute induced velocity values in the wake (in hover) are on the order of 11 m/s.
\section{NURBS-based Vortex Filament Method}%
    Although parametric equations would be sufficient to model the vortex rings as used in this paper, a more flexible approach using Non-Uniform Rational B-Spline (NURBS) curves (Ref.~\cite{piegl_tiller_book_1997}) was adopted for this research. For elementary cases such as circles, an exact geometry representation is available using rational polynomials. When this is not possible, accurate approximations can be constructed easily (see e.g. Ref.~\cite{hoydonck_gerritsma_tooren_2011}). In most filament methods, analytical solutions of the Biot-Savart law for basic elements of the wake are used. In this paper, induced velocity values are computed by numerical integration of the regularised Biot-Savart law instead (Eq.~\ref{eq:mollified_biot_savart_param_eqn}). Then, adaptive quadrature methods can be used to compute values to required accuracy. Or, without changing the resolution of the filament geometry, it is possible to trade computing speed for accuracy. One disadvantage of the current method is that it can only be used when the three-dimensional smoothing associated with a certain swirl velocity profile is known (see Table~\ref{tab:velocity_smoothings}).
    
    \subsection{Parametric Geometry Representation}
    In this section, some background information is presented related to Non-Uniform Rational B-Splines. The focus is on concepts that are necessary for a practical implementation. In-depth information on the mathematical background of NURBS can be found in the book by Piegl and Tiller (Ref.~\cite{piegl_tiller_book_1997}). Topics discussed here are the constituent parts of NURBS curves: knot vectors and basis functions. The equations to calculate point positions and derivative vectors on NURBS curves will be shown and finally, a method to construction NURBS circles is discussed.
    \subsubsection{Knot Vectors}%
        As generalisations of B\'{e}zier curves, (non-rational) B-spline and (rational) NURBS curves are constructed by joining together multiple polynomial segments of a desired degree to form a single curve. The breakpoints in parametric space along the curve where the individual polynomials are joined are called the \emph{knots}. The array of non-decreasing parameters that separate the segments is called the \emph{knot vector} $\mb{U}$ and has the form
        $\mb{U} = [u_0,u_1,\ldots,u_{m}]$ with $u_i \leq u_{i+1}, i = 0,\ldots,m-1$.
        
        A \emph{uniform} knot vector has equally-spaced knots. In \emph{normalised} form, the first knot has value 0 and the last knot equals 1. To ensure that a curve interpolates the end points of the control polygon over which it is defined, the knot vector is \emph{clamped}: the multiplicity of the first and last values in the knot vector is increased to $p+1$,
        \begin{equation}\label{eq:clamped_knot_vec}
            \mb{U} = [\underbrace{a,\ldots,a}_{p+1},u_{p+1},\ldots,u_{m-p-1},\underbrace{b,\ldots,b}_{p+1}],
        \end{equation}
        where $p$ is the degree of the curve. %For a $p^{th}$ degree curve, a clamped knot vector with $m+1$ knots and end multiplicities of $p+1$ requires $n+1$ control points, $m = n + p + 1$.
    \subsubsection{Basis Functions}%
        The $i$\textsuperscript{th} B-spline basis function of degree $p$, denoted by $N_{i,p}(u)$, is defined recursively by
        \begin{equation}\label{eq:bspline_basis_function_definition}
            \begin{split}
            N_{i,0}(u) &= \begin{cases}
                        1& \textrm{if}\quad u_i \leq u < u_{i+1} \\
                        0& \textrm{otherwise}
                        \end{cases}\\
            N_{i,p}(u) &= \frac{u-u_i}{u_{i+p}-u_i}N_{i,p-1}(u) + \frac{u_{i+p+1} - u}{u_{i+p+1}-u_{i+1}} N_{i+1,p-1}(u).
            \end{split}
        \end{equation}
        When the Biot-Savart law is used to compute the induced velocity at a point due to a vortex filament, first derivative vectors are needed at quadrature points on the filament. These derivatives can be expressed in terms of the first derivatives of the basis functions of the curves, which in turn are defined as
        \begin{equation}\label{eq:bspline_basis_fun_first_deriv}
            N_{i,p}^{'} = \frac{p}{u_{i+p}-u_{i}}N_{i,p-1}(u) - \frac{p}{u_{i+p+1}-i_{i+1}} N_{i+1,p-1}(u).
        \end{equation}
    \subsubsection{Calculation of Point Locations and First Derivatives on NURBS Curves}
        A $p$\textsuperscript{th} degree NURBS curve is defined by
        \begin{equation}\label{eq:nurbs_curve_def}
            \mb{C}(u) = \left(\sum_{i=0}^{n} N_{i,p}(u) w_i \mb{P}_i\right) \Bigg/ \left(\sum_{i=0}^{n} N_{i,p}(u) w_i\right) \qquad a \leq u \leq b,
        \end{equation}
        with $\mb{P}_i$ the control points and $w_i$ the weights of the control points. $N_{i,p}(u)$ are the $p$\textsuperscript{th}-degree B-spline basis functions defined on the non-periodic and non-uniform knot vector $\mb{U}$ (Eq.~\ref{eq:clamped_knot_vec}).
        
        Equation~\ref{eq:nurbs_curve_def} can be written as a non-rational, piecewise polynomial curve using rational basis functions $R_{i,p}(u)$,
         \begin{equation}\label{eq:nurbs_curve_def_rat_basis}
            \mb{C}(u) = \sum_{i=0}^{n} R_{i,p}(u) \mb{P}_i
        \end{equation}
        where $R_{i,p}(u) = \frac{N_{i,p}(u) w_i}{\sum_{j=0}^{n}N_{j,p}(u)w_j}$.
        
        In the non-rational case all weights $w_i$ have unit value and Eq.~\ref{eq:nurbs_curve_def} reduces to
        \begin{equation}\label{eq:bspline_curve_def}
            \mb{C}(u) = \sum_{i=0}^{n} N_{i,p}(u) \mb{P}_i \qquad a \leq u \leq b.
        \end{equation}
        The first derivative at a point $u$ along a polynomial B-spline curve is given by
        \begin{equation}\label{eq:bspline_curve_deriv}
            \mb{C}'(u) = \sum_{i=0}^{n} N_{i,p}'(u) \mb{P}_i,
        \end{equation}
        where the first derivative of the basis function $N_{i,p}'(u)$ is given by Eq.~\ref{eq:bspline_basis_fun_first_deriv}. 
        
        Using \emph{homogeneous coordinates}, NURBS curves can be represented in a compact and efficient manner as non-rational, piecewise polynomial B-spline curves in four-dimensional space using weighted control points $\mb{P}_i^w = (w_ix_i,w_iy_i,w_iz_i,w_i)$,
        \begin{equation}
            \mb{C}^w(u) = \sum_{i=0}^{n} N_{i,p}(u) \mb{P}_i^w.
        \end{equation}
        For a (rational) NURBS curve, Eq.~\ref{eq:bspline_curve_deriv} can be used directly to calculate the derivatives at a point along the curve in homogeneous coordinates
        \begin{equation}\label{eq:nurbs_curve_deriv_homogeneous_coords}
            \mb{C}'^w(u) = \sum_{i=0}^{n} N_{i,p}'(u) \mb{P}^{w}_i.
        \end{equation}
        Setting $\mb{C}(u) = \frac{w(u)\mb{C}(u)}{w(u)} = \frac{\mb{A}(u)}{w(u)}$, where $\mb{A}(u)$ is the vector-valued function whose coordinates are the first three coordinates of $\mb{C}^w(u)$. Then (Ref.~\cite{piegl_tiller_book_1997}, p. 125),
        \begin{equation}\label{eq:nurbs_curve_first_deriv}
            \mb{C}'(u) = \frac{\mb{A}'(u) - w'(u)\mb{C}(u)}{w(u)},
        \end{equation}
        where $\mb{A}'(u)$ is the vector-valued function whose coordinates are the first three coordinates of $\mb{C}'^w(u)$ and $w'(u)$ is its last coordinate.
    \subsubsection{Construction of Circles}
\begin{figure}
    \centering
    \includegraphics{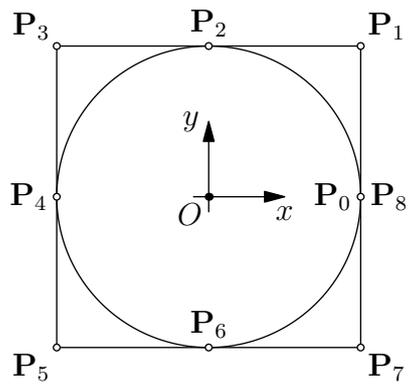}
    \caption{Geometry of the NURBS circle as used to compute vortex ring induced velocities.}\label{fig:nurbs_geometry_reference}
\end{figure}

        Multiple methods can be used to construct circles using NURBS curves (see Ref.~\cite{piegl_tiller_1989}). One of the most straighforward approaches uses a set of rational quadratic B\'{e}zier arcs pieced together by a knot vector with double internal knots.
        As an example, a circle constructed from four rational B\'{e}zier segments is shown in Fig.~\ref{fig:nurbs_geometry_reference}.
        The weights for the even-numbered control points are one and the weights for the odd-numbered control points can be found to be $w_i = \cos \theta$,
        where $\theta$ is the angle formed by the triplets $\angle \mb{P}_{i}\mb{P}_{i-1}\mb{P}_{i+1}$ with $\mb{P}_i$ one of the odd-numbered control points. In case of a four-segment (9-point) circle (Fig.~\ref{fig:nurbs_geometry_reference}), this angle is 45\textdegree, so that the weights of the odd-numbered control points equal $\frac{\sqrt{2}}{2} \approx 0.7071$.
    \subsection{Static Validation}%
        The numerical method is validated by comparing calculated induced velocity values at a point on a vortex ring of unit strength with the analytical results derived in the previous section. For a range of relative core sizes $\sigma/R$ between $10^{-6}$ and 1, the induced velocity is computed at the point $(-R,0)$. The geometry of the NURBS-based vortex ring is depicted in Fig.~\ref{fig:nurbs_geometry_reference}.
        
        The adaptive quadrature method developed by Shampine (Ref.~\cite{shampine_2008}) was adapted for induced velocity computations (the original method only works for scalar-valued functions, while induced velocity computations are vector-valued). 
        The relative and absolute errors were set to $10^{-5}$ and $10^{-10}$, respectively. The maximum number of subdivision levels is ten and upon subdivision, segments are divided in four subsegments.
        
        Induced velocity values are computed for the core with solid body rotation (Eq.~\ref{eq:vr_numerical_rankine}), the Rosenhead-Moore kernel (Eq.~\ref{eq:vr_rosenhead_moore}) and lastly, for the Ramasamy-Leishman kernel at vortex Reynolds numbers of $10^4$ and $10^6$ (Eq.~\ref{eq:vr_numerical_ramasamy_leishman}).
        
\begin{figure}
    \centering
    \subfloat[][]{\label{subfig:vrv_smoothings_abs}\includegraphics{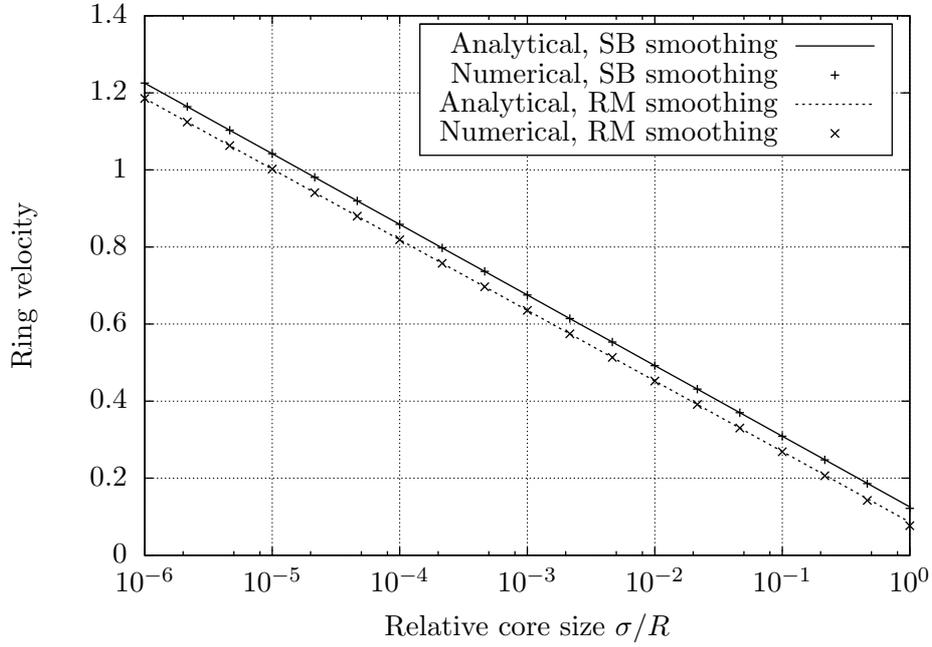}}\hspace{.05\textwidth}%
    \subfloat[][]{\label{subfig:vrv_smoothings_rel}\includegraphics{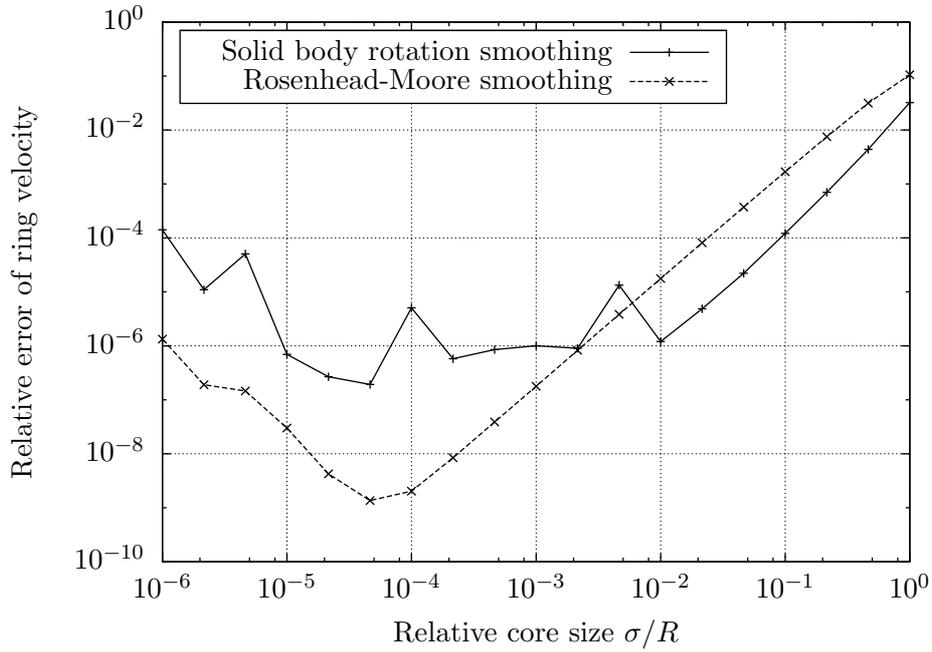}}%
    \caption{Viscous vortex ring velocity as a function of relative core size $\sigma/R$ for the smoothings of Rosenhead-Moore and Rankine,~\protect\subref{subfig:vrv_smoothings_abs} absolute values and~\protect\subref{subfig:vrv_smoothings_rel} relative errors.}\label{fig:vrv_smoothings_1}
\end{figure}

\begin{figure}
    \centering
    \subfloat[][]{\label{subfig:vrv_smoothings_rl_abs}\includegraphics{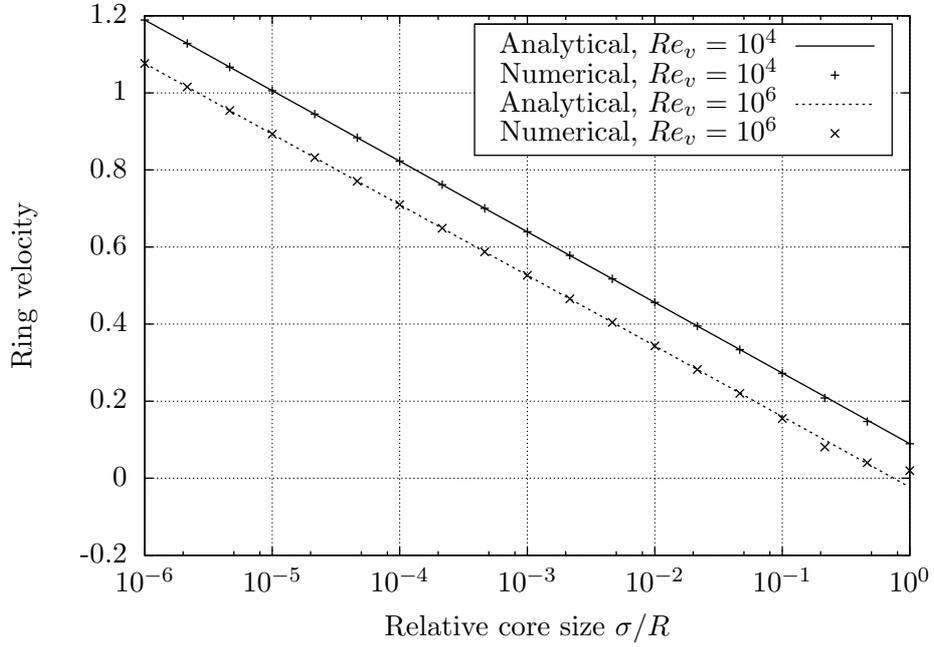}}\hspace{.05\textwidth}%
    \subfloat[][]{\label{subfig:vrv_smoothings_rl_rel}\includegraphics{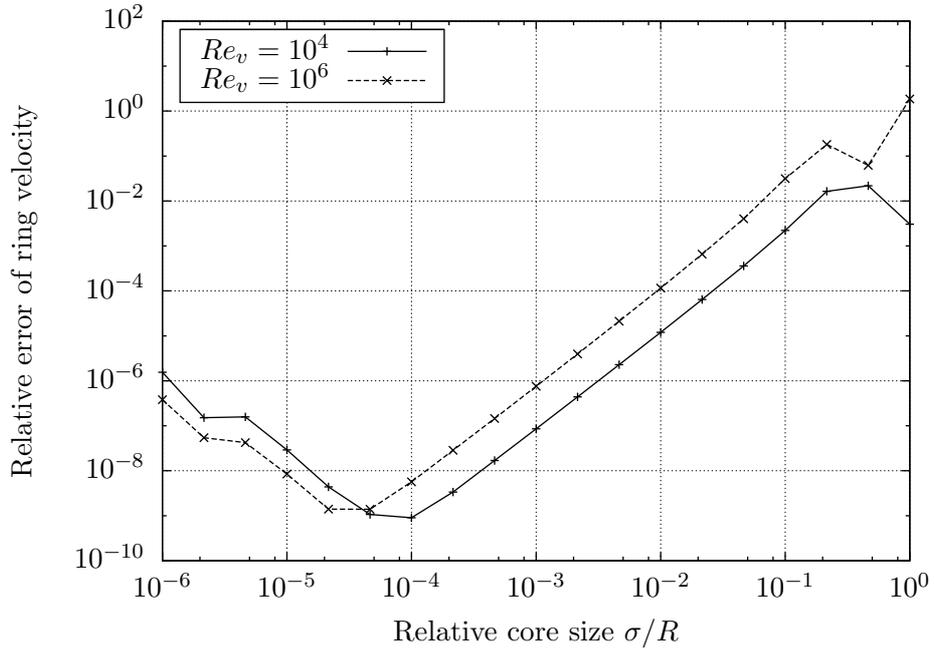}}%
    \caption{Viscous vortex ring velocity as a function of relative core size $\sigma/R$ for the smoothing based on the Ramasamy-Leishman swirl velocity profile at vortex Reynolds numbers of $10^4$ and $10^6$,~\protect\subref{subfig:vrv_smoothings_rl_abs} absolute values and~\protect\subref{subfig:vrv_smoothings_rl_rel} relative errors.}\label{fig:vrv_smoothings_2}
\end{figure}

        Inspection of Figs.~\ref{fig:vrv_smoothings_1} and~\ref{fig:vrv_smoothings_2} shows that the numerical and analytical results converge towards each other as the relative core size is decreased with a trend that is approximately second-order. For relative core sizes smaller than $10^{-4}$, the second-order trend breaks down. For very small core sizes, the function that is integrated starts to resemble finite pulses with a very narrow base.
        The velocity smoothing associated with solid body rotation is the only one that is only $C^0$ continuous. As a result, the function integrated is not smooth and the numerical result breaks down earlier (Fig.~\ref{subfig:vrv_smoothings_rel}). These are two examples that are difficult for all numerical quadrature methods.

\section{Analysis of Error Sources}
    For the segmentation method, two sources of error have been identified in the past (Ref.~\cite{bliss_teske_quackenbush_1987}). These are errors caused by the incorrect position of straight vortex segments (location errors) and errors due to neglecting the local curvature contributions to induced velocity values (curvature errors). An additional source of error (which was first identified in Ref.~\cite{hoydonck_tooren_2011}) is due to the use of the core correction model as introduced by Scully (Ref.~\cite{scully_1975}). First, the impact of this correction on the accuracy and convergence characteristics of induced velocity computations will be analysed. Afterwards, the importance of local curvature corrections on induced velocity computations will be investigated. In both cases, the NURBS-based vortex filament method is used in the analysis.
    \subsection{Influence of Unconditional Use of Perpendicular Distance on Accuracy}
        The unconditional use of the perpendicular distance in core correction models has a detrimental effect on the accuracy of computed velocity of a viscous vortex ring. Using the NURBS-based vortex filament method, the induced velocity near a vortex ring with the normalised Gaussian velocity smoothing is computed, both with the correct smoothing,
        \begin{equation}\label{eq:gauss_vel_smoothing}
            g_3(\rho) = \frac{\textrm{erf}(\rho\sqrt{a}) - 2\rho\sqrt{\frac{a}{\pi}} e^{-a \rho^2} }{4\pi}
        \end{equation}
        as given in Table~\ref{tab:velocity_smoothings} and its equivalent that uses the perpendicular distance,
        \begin{equation}\label{eq:gauss_vel_smoothing_perp_dist}
            g_3(p) = \frac{\textrm{erf}(p\sqrt{a}) - 2p\sqrt{\frac{a}{\pi}} e^{-a p^2} }{4\pi},
        \end{equation}
        where $p = h/\sigma$ is the perpendicular distance divided by the core size.
        
        The velocity induced by the viscous vortex ring is computed in the plane of the ring (where only the vertical component is nonzero) on a square with sides ten times the ring radius using quadrature rules with eight or sixteen abscissae (per quadrant, one rule is used, so induced velocity values are computed with either $4\times8=32$ or $4\times16=64$ evaluations of the Biot-Savart law). Results are compared with the analytical solution for a potential vortex ring (Ref.~\cite{castles_leeuw_1954}). Close to the potential vortex ring, the magnitude of the induced velocity increases without bounds. When these results are used as a reference for the numerical values computed with the viscous ring, it is expected that the errors will be large for points close to the ring. For evaluations points in the far field, the numerical results should converge to the analytical values.
        
\begin{figure}
    \centering
    \subfloat[][]{\label{subfig:ind_vel_rel_err_2d}\includegraphics[scale=0.7]{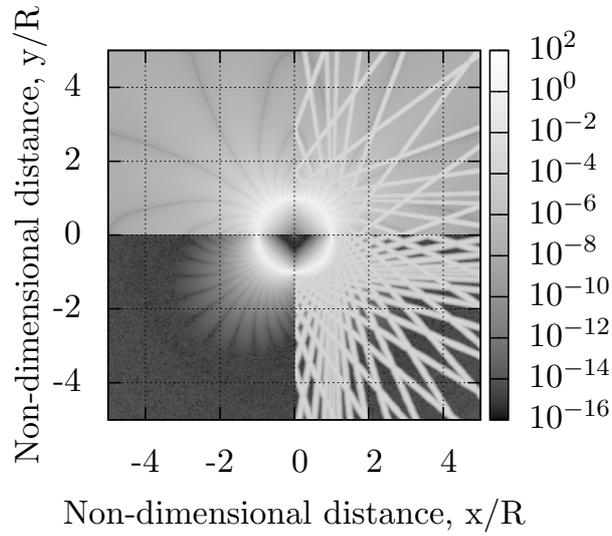}}\\
    \subfloat[][]{\label{subfig:ind_vel_rel_err_1d}\includegraphics{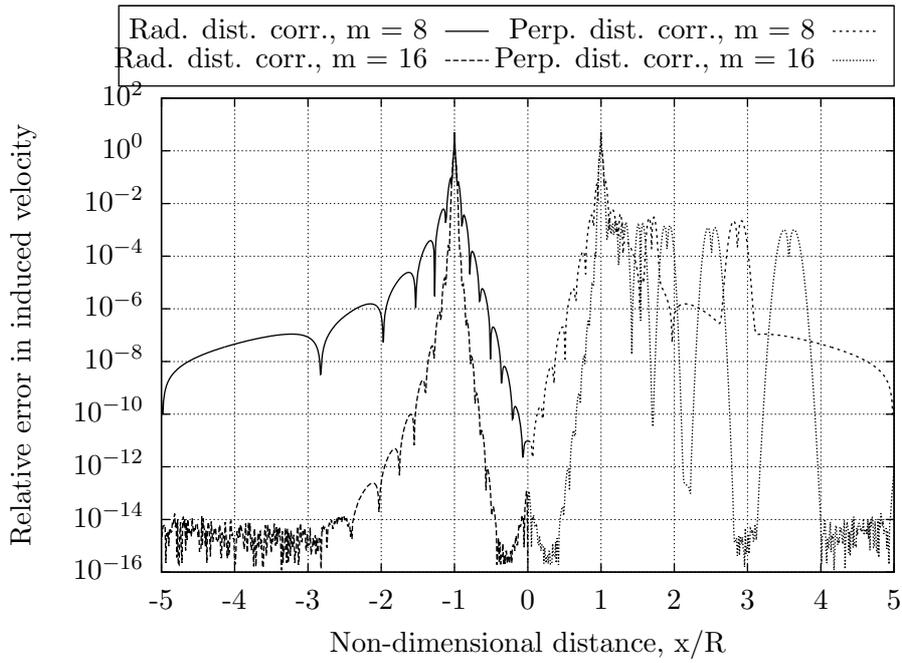}}
    \caption{Convergence of induced velocity values computed in the plane of a viscous vortex ring with the normalised Gaussian smoothing using the NURBS-based vortex filament method. Results on the right-hand side are computed with the perpendicular core smoothing, while results on the left-hand side are computed with the correct, radial smoothing. In~\protect\subref{subfig:ind_vel_rel_err_2d}, results for quadrature rules with 8 abscissae are shown in the top half while those for 16 abscissae are shown in the bottom half. A cross section of the data along the x-axis is shown in~\protect\subref{subfig:ind_vel_rel_err_1d}.}\label{fig:ind_vel_rel_err_1d_and_2d}
\end{figure}

        The left-hand side of Fig.~\ref{subfig:ind_vel_rel_err_2d} shows the relative errors w.r.t. the analytical results when Eq.~\ref{eq:gauss_vel_smoothing} is used. The right-hand side of this figure show the relative errors when Eq.~\ref{eq:gauss_vel_smoothing_perp_dist} is used instead. The upper side shows results for $m=8$, the bottom side shows results for $m=16$. A cross section of this data along the x-axis is shown in Fig.~\ref{subfig:ind_vel_rel_err_1d}. Again, the left-hand shows results computed with Eq.~\ref{eq:gauss_vel_smoothing} and the right hand side of the figure shows results for Eq.~\ref{eq:gauss_vel_smoothing_perp_dist}. 
        
        As expected, convergence towards the analytical results is slow near the vortex ring. Still, doubling the number of abscissae of the quadrature rule gives results with approximately twice as many correct digits. These figures clearly show that when Eq.~\ref{eq:gauss_vel_smoothing_perp_dist} is used as smoothing function, results do not converge to the analytical values for evaluation points located on or near tangent lines of the quadrature points on the vortex ring. Doubling the number of abscissae has the exact opposite effect, for more points in the plane of the ring, results fail to converge to the correct values. The width of the regions where results do not converge is related to the convergence characteristics of the smoothing used. For example, the width for the Rankine solid body rotation model is exactly twice the core size, for other smoothings, it is larger. When one would focus on comparing results for evaluation points located on the inside of the vortex ring, these errors would not show up at all. 
        
\begin{figure}
    \centering
    \subfloat[][]{\label{subfig:biot_savart_function_form}\includegraphics{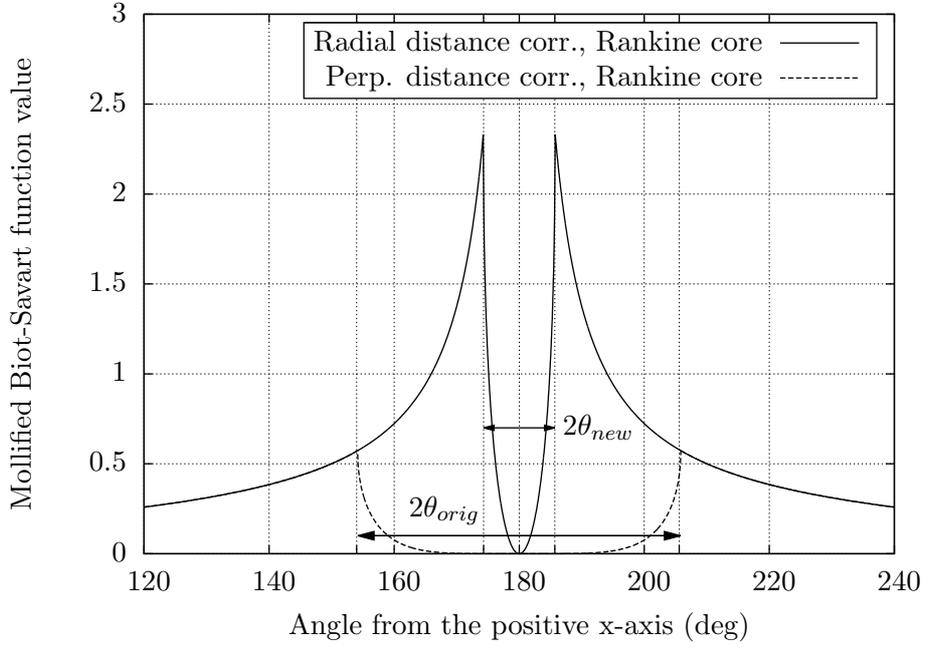}}\\
    \subfloat[][]{\label{subfig:core_model_correction_sectors}\includegraphics{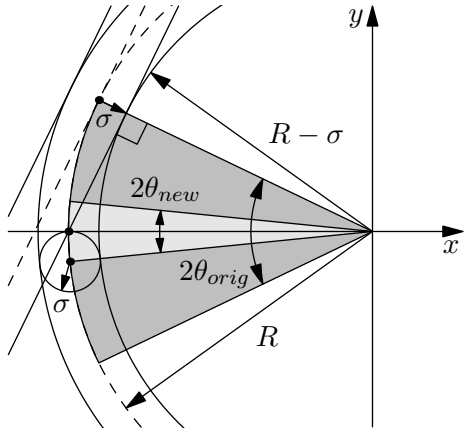}}
    \caption{Boundaries between potential and corrected (viscous) Biot-Savart function values for a point $(-R,0)$ on a vortex ring ($\sigma/R = 0.1$) for the Rankine swirl velocity profile. For both corrections, the area under the curve from 0\textdegree~to 360\textdegree~ in~\protect\subref{subfig:biot_savart_function_form} equals the induced velocity at the evaluation point due to the whole vortex ring. The width of the correction regions are $\theta_{orig} = 25.8^{\circ}$ (using Eq.~\protect\ref{eq:perp_dist_arc_length}) and $\theta_{new} = 5.7^{\circ}$ (using Eq.~\protect\ref{eq:rad_dist_arc_length}).}\label{fig:rankine_potential_viscous_boundaries}
\end{figure}

        When the velocity on the ring itself is computed, similar results are obtained, the computed value is reduced with respect to the reference value (Ref.~\cite{hoydonck_tooren_2011}).
        This is illustrated in Fig.~\ref{fig:rankine_potential_viscous_boundaries} where smoothed Biot-Savart function values are shown as a function of the azimuth angle of a quadrature point on the vortex ring. The area under the curve from 0 to $2\pi$ equals the value of the induced velocity at the evaluation point $(-R,0)$. In this case, the solid body rotation velocity smoothing is used with a relative core size of 10\%. This particular smoothing gives a sharp boundary between smoothed and unsmoothed function values. It is clear from this figure that the unconditional use of the perpendicular distance to correct the Biot-Savart law results in a larger area where the potential value is corrected (or reduced) and consequently, gives a value for the induced velocity of the ring that is lower than the correct one.
        
        Using Fig.~\ref{fig:rankine_potential_viscous_boundaries}, the arc length of the smoothing region as a function of the relative core size is computed. For the perpendicular smoothing, the width of the region is proportional to the square root of twice the relative core size,
        \begin{equation}\label{eq:perp_dist_arc_length}
            \theta_{orig} =  \cos^{-1}(1-\sigma/R) \approx \sqrt{2\sigma/R} \gg \sigma/R \quad \textrm{for} \quad \sigma \ll R.
        \end{equation}
        When the radial distance is used, the width of the region is proportional to the relative core size,
        \begin{equation}\label{eq:rad_dist_arc_length}
            \theta_{new} =  2 \sin^{-1}(\sigma/(2R)) \approx \sigma/R,
        \end{equation}
        which confirms the conclusion drawn from Fig.~\ref{fig:rankine_potential_viscous_boundaries}: the width of the smoothing region must be wider for the perpendicular smoothing than for the radial smoothing and therefore the computed induced velocity computed with the former smoothing is always lower than the one computed with the latter smoothing.
    \subsection{Influence of Local Curvature on Induced Velocity}%
        In methods where curved filaments are discretised using straight segments, the two segments directly adjacent to a marker point do not contribute anything to the induced velocity at that point. This contribution can be taken into account by fitting a circular arc throught the points and using an approximate analytical expression for the velocity induced by the arc (e.g. Refs.~\cite{scully_1975} and~\cite{leonard_1975}). Using the NURBS-based filament method, the influence of a filament arc at its midpoint will be computed for a range of arc lengths and relative core sizes, for various smoothings.
        
\begin{figure}
    \centering
    \includegraphics[]{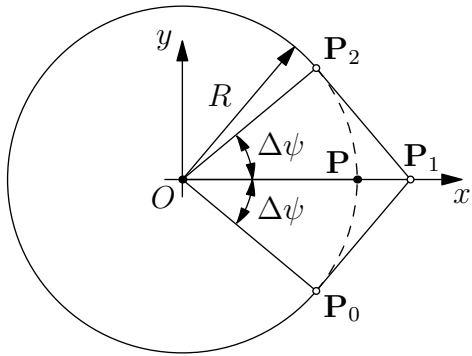}
    \caption{Configuration of the portion of the vortex ring that is not taken into account when computing induced velocities at point $\mb{P}$. The induced velocity is computed from the solid part of the circle by substracting the local contribution from the velocity of the complete vortex ring.}\label{fig:nurbs_arc_iv_at_mid_point}
\end{figure}

        The filament configuration is shown in Fig.~\ref{fig:nurbs_arc_iv_at_mid_point}. The circular arc with angle $2\Delta\psi$ is represented exactly with a quadratic, three-point, rational NURBS curve. The control points are labelled $\mb{P}_0$, $\mb{P}_1$ and $\mb{P}_2$. Induced velocity is evaluated halfway the curve, where it crosses the positive x-axis. Since the curve is completely located in the xy-plane, the induced velocity at the evaluation point only has a non-zero z-component.
        
\begin{figure}
    \centering
    \includegraphics{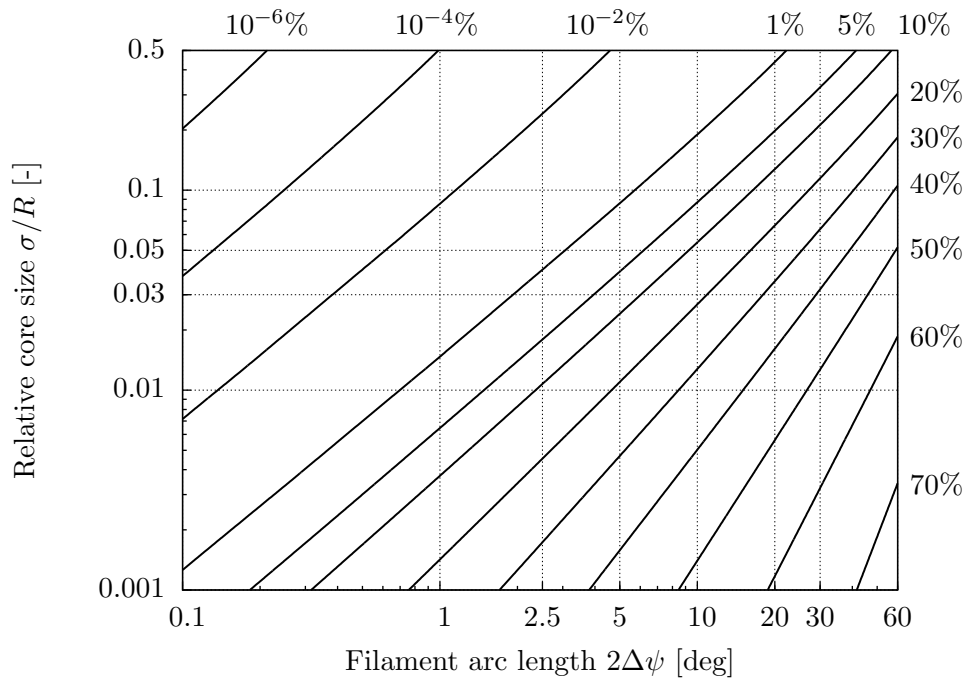}
    \caption{Relative error in predicted velocity at a point on a vortex ring with the Rosenhead-Moore velocity smoothing by neglecting the influence of the contribution of the adjacent part of the ring.}\label{fig:rel_vel_arc_rosenhead_moore}
\end{figure}

\begin{figure}
    \centering
    \includegraphics{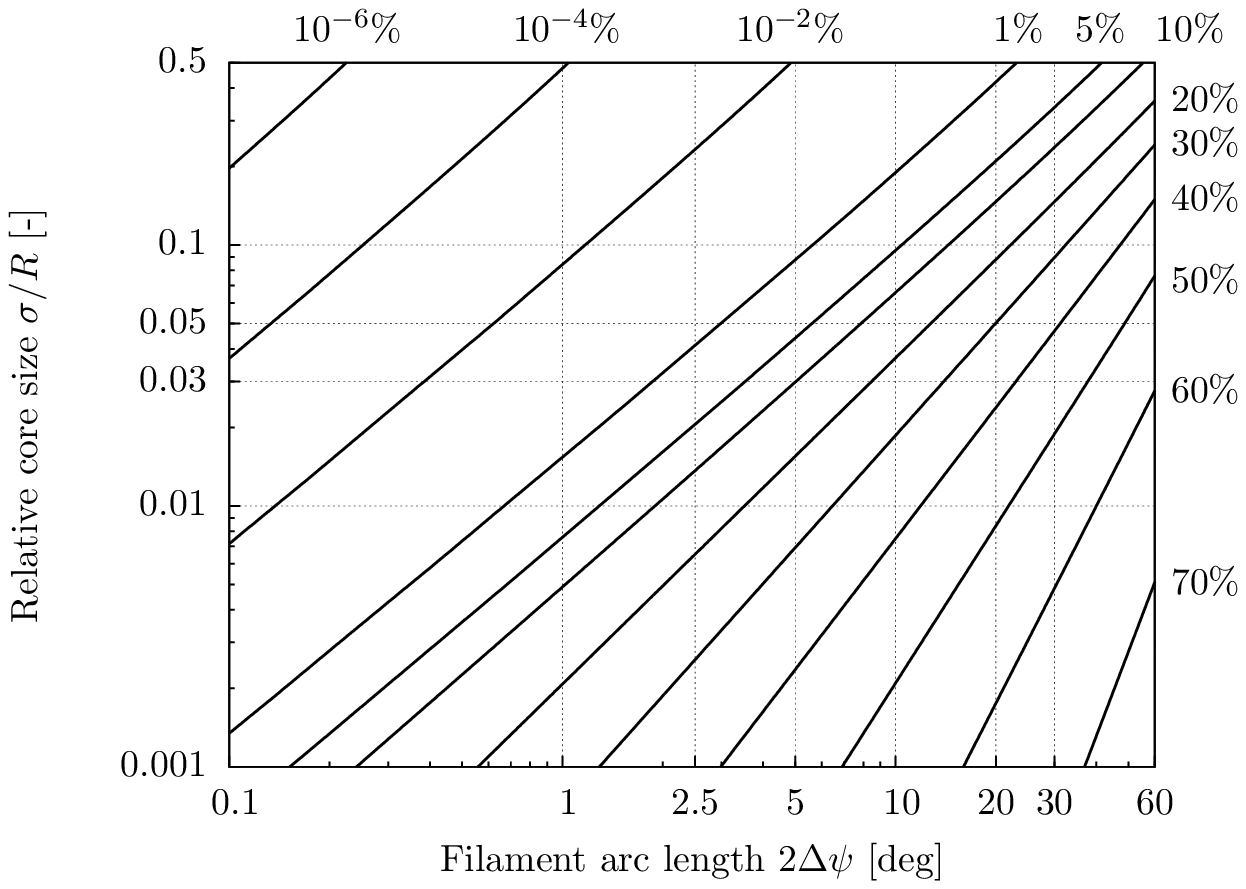}
    \caption{Relative error in predicted velocity at a point on a vortex ring with the normalised Gaussian velocity smoothing by neglecting the influence of the contribution of the adjacent part of the ring.}\label{fig:rel_vel_arc_normalised_gaussian}
\end{figure}

        For smoothings that are at least $C^1$ continuous, results are computed for relative core sizes between 0.001 and 0.5 and for filament arc lengths in the range 0.1 to 60 degrees. The relative errors between the contribution due to the adjacent segment and the velocity due to the whole vortex ring are shown in Figs.~\ref{fig:rel_vel_arc_rosenhead_moore} and~\ref{fig:rel_vel_arc_normalised_gaussian} for the Rosenhead-Moore and normalised Gaussian correction.
        In these figures, the relative errors are computed as the absolute difference between the contribution due to the adjacent segment and the velocity due to the whole vortex ring divided by the value of the ring velocity,
        $\Delta v_z = \frac{\textrm{abs}(v_{z,arc} - v_{z,ring})}{\textrm{abs}(v_{z,ring})}$.

        Inspection of Figs.~\ref{fig:rel_vel_arc_rosenhead_moore} and~\ref{fig:rel_vel_arc_normalised_gaussian} shows the same general trends, the relative magnitude of the local contribution to the induced velocity at a point on a vortex ring is larger for increasingly bigger arc lengths and for smaller core sizes. At discretisation resolutions as used in the first generation of wake codes ($\Delta\psi \approx 15^{\circ}$), neglecting the local contribution results in serious underpredictions of the induced velocity. For example, an error of approximately 40\% is made for a vortex ring with a relative core size of 3\% for the Rosenhead-Moore smoothing (Fig.~\ref{fig:rel_vel_arc_rosenhead_moore}) when the local curvature contribution is neglected. At a fixed discretisation level and relative core size, the relative error becomes smaller for increasing vortex Reynolds numbers. For example, at a discretisation of $\Delta\psi=2.5^{\circ}$, the error is 10 percent for a vortex Reynolds number of one (Fig.~\ref{fig:rel_vel_arc_normalised_gaussian}). At a vortex Reynolds number of $10^5$, the error is approximately five percent and for a vortex Reynolds number of one million, it is even smaller.
        
\begin{figure}
    \centering
    \includegraphics{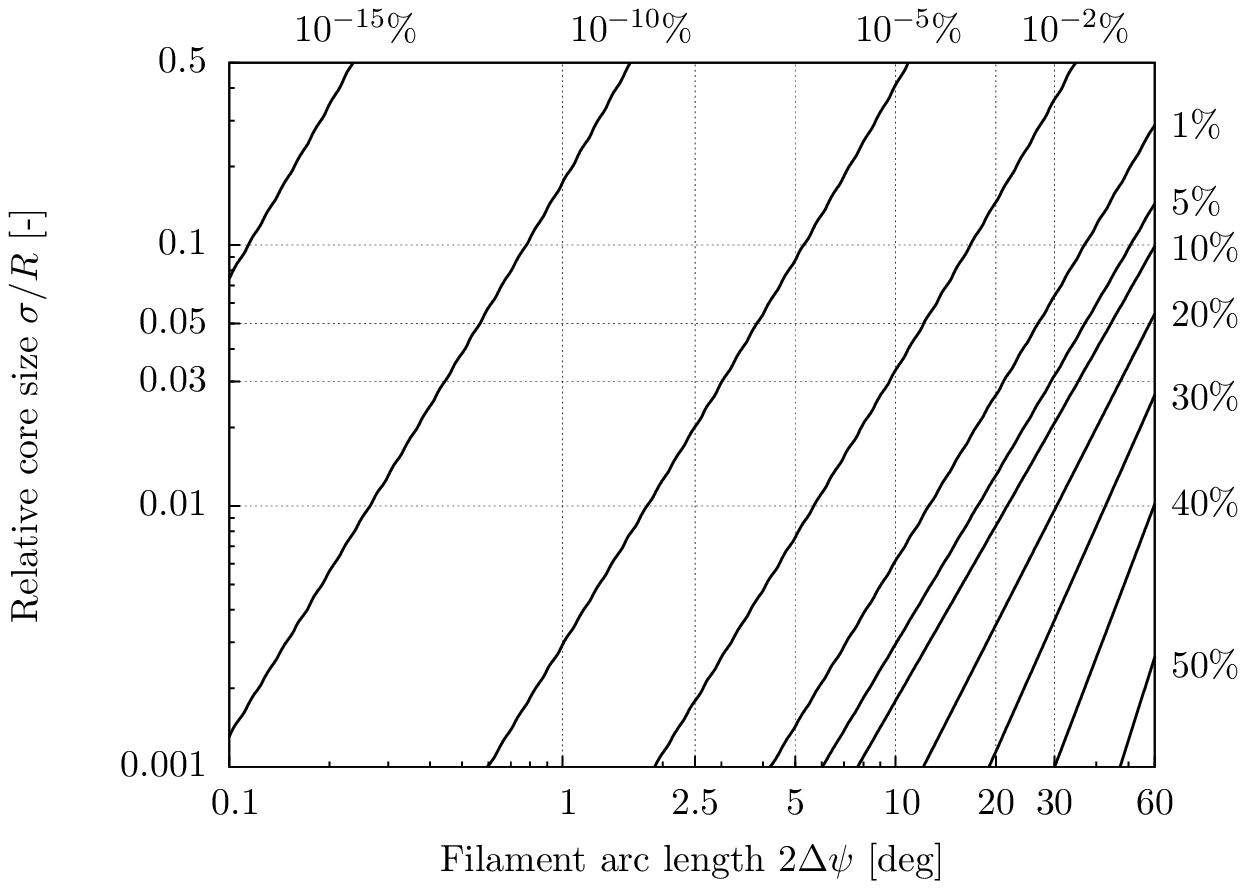}
    \caption{Relative error in predicted velocity at a point on a vortex ring with the normalised Gaussian velocity smoothing by neglecting the influence of the contribution of the adjacent part of the ring for the perpendicular smoothing, using Eq.~\protect\ref{eq:gauss_vel_smoothing_perp_dist}.}\label{fig:rel_vel_arc_normalised_gaussian_old_corr}
\end{figure}

        Results shown here were computed with the correct (radial) smoothing, not with smoothings that use the perpendicular distance, as most methods do (Ref.~\cite{scully_1975,bagai_phd_1995}). It follows that the error contours shown in Figs.~\ref{fig:rel_vel_arc_rosenhead_moore} and~\ref{fig:rel_vel_arc_normalised_gaussian} cannot be used to get an estimate of the error for segmentation methods that use Scully's core correction method. When the perpendicular smoothing is used (using for example Eq.~\ref{eq:gauss_vel_smoothing_perp_dist} instead of Eq.~\ref{eq:gauss_vel_smoothing}), errors are orders smaller at fine discretisations (compare results in Fig.~\ref{fig:rel_vel_arc_normalised_gaussian} with those in Fig.~\ref{fig:rel_vel_arc_normalised_gaussian_old_corr}). For example, at a relative core size of 0.03, the one percent error line shift from $\Delta\psi\approx 1^{\circ}$ right to $\Delta\psi\approx 10^{\circ}$. From this result, one could draw erronous conclusions regarding the need for curvature corrections in a segmentation method.

\section{Convergence Characteristics of the Segmentation Method}%
    In this final section, the convergence characteristics of the segmentation method will be analysed. The velocity induced by a viscous vortex ring with a certain smoothing is computed at three different locations, the first one on the ring at $(R,0)$, the second one in the centre of the ring and the third one on the outside of the ring, at $(\frac{6}{5}R,0)$. In a free wake model, velocities are mainly computed at marker points on the vortex filaments. The results presented in this section for the point located on the ring itself give a realistic representation of the convergence characteristics for these marker points, while the last two points are representative for far-field computations. An alternative would be to evaluate velocities at a large set of points near the vortex ring for various discretisation resolutions, and per discretisation, compute the norm of the error. This way unfavourable convergence at a small set of points may be hidden when convergence at most points is good. To avoid this, results will be shown per point separately. 
    
    The vortex ring is approximated with a set of $n$ straight segments of arc length $\frac{2\pi}{n}$. The two modeling options analysed in the previous section are combined in four separate cases: 1) original core correction without curvature correction, 2) original core correction with curvature correction, 3) improved core correction without curvature correction and 4) improved core correction with curvature correction.
    
    The curvature corrections are computed with the NURBS-based filament method. Per smoothing, induced velocity values are computed halfway on a segment of a circular arc for a range of arc lengths and relative core sizes. By storing these values in a two-dimensional array, a simple table lookup can be used to fetch the correct value. In more realistic cases, the exact core sizes and arc lengths are not known in advance and two-dimensional interpolation should be used.
    
\begin{figure}
    \centering
    \subfloat[][Original core correction.]{\label{subfig:original_straight_segm_core_correction}\includegraphics{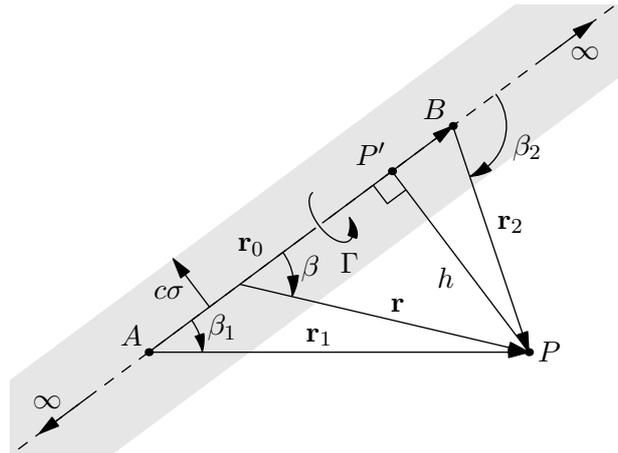}}\hspace{1cm}
    \subfloat[][Improved core correction.]{\label{subfig:improved_straight_segm_core_correction}\includegraphics{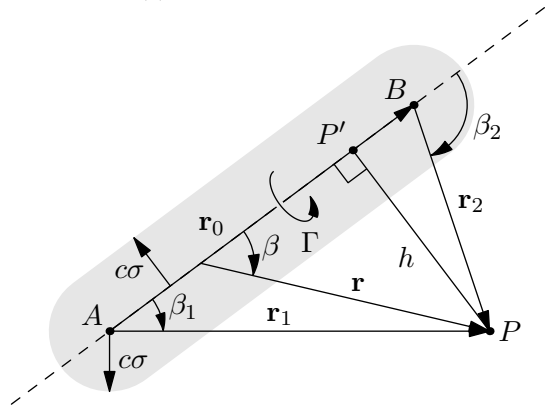}}
    \caption{Two core corrections that can be used to correct the velocity field induced by a finite, straight vortex line segment. The gray region is the area where the correction is applied. Its width depends on the convergence characteristics of the specific smoothing, for the solid body rotation smoothing, $c=1$, for other smoothings, $c \gg 1$.}\label{fig:straight_segm_core_corrections}
\end{figure}

    The original core correction uses the perpendicular distance to an evaluation point (Fig.~\ref{subfig:original_straight_segm_core_correction}) to correct the analytical formula for the induced velocity. For the improved core correction proposed in Ref.~\cite{hoydonck_tooren_2011}, the correction factor depends on the location of the evaluation point w.r.t. the endpoints of the segment (see Fig.~\ref{subfig:improved_straight_segm_core_correction}). When its projection falls between the end points, the perpendicular distance is used as before. If the projection point lies beyond one of the end points, the radial distance to the respective end point is used. This way, corrections are restricted to a finite area close to the segment.
    
    For the point located on the vortex ring, the analytical formulas for the induced velocity of a viscous ring with a circular core derived earlier are used as reference. For the other two points, induced velocity values are compared with values computed for a potential vortex ring (Ref.~\cite{castles_leeuw_1954}).
    
    For each smoothing studied, the relative core size $\sigma/R$ was varied between $0.001$ and $0.5$ and the number of segments was varied between 5 ($\Delta\psi=72^{\circ}$) and 3600 ($\Delta\psi=0.1^{\circ}$). This covers the complete range of tip vortex core sizes and filament discretisations that is used in engineering rotorcraft applications. A small, but representative set of these results will be shown and discussed hereafter.
    
    For the point located in the centre of the vortex ring, no curvature corrections are applied. In addition, the point is located on the concave side of the vortex ring where the use of the improved core correction makes no difference over the orginal one. The results for the four different cases all collapse onto a single line with a second-order slope.
    
\begin{figure}
    \centering
    \includegraphics{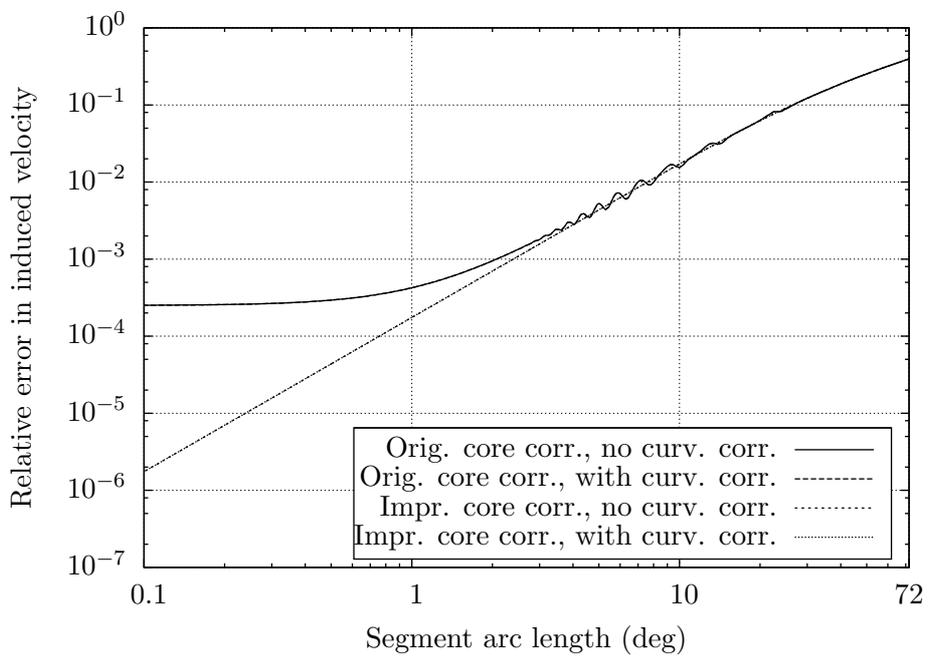}
    \caption{Convergence of straight-line segmentation method at $(0,1.2R)$ for a viscous vortex ring with Gaussian smoothing ($\sigma/R = 0.03$). Convergence is not satisfactory for the original core correction model.}\label{fig:segm_conv_outside_point_relerr_gaussian}
\end{figure}

    Figure~\ref{fig:segm_conv_outside_point_relerr_gaussian} shows results for the point on the outside of the vortex ring (at $(\frac{6}{5}R,0)$). Similar to the case where the evaluation point is located on the inside of the ring, curvature corrections are zero. Recalling the results from the previous section (Fig.~\ref{subfig:ind_vel_rel_err_2d}), results may not converge completely to the correct value for the original core correction. This is indeed the case: at coarse discretisations, second-order convergence is indeed achieved, but for segment arc lengths smaller than one degree, convergence stops completely. For the improved core correction, convergence is not impeded, even at the finest discretisations.
    
\begin{figure}
    \centering
    \subfloat[][]{\label{subfig:segm_conv_ring_point_absval_gaussian}\includegraphics{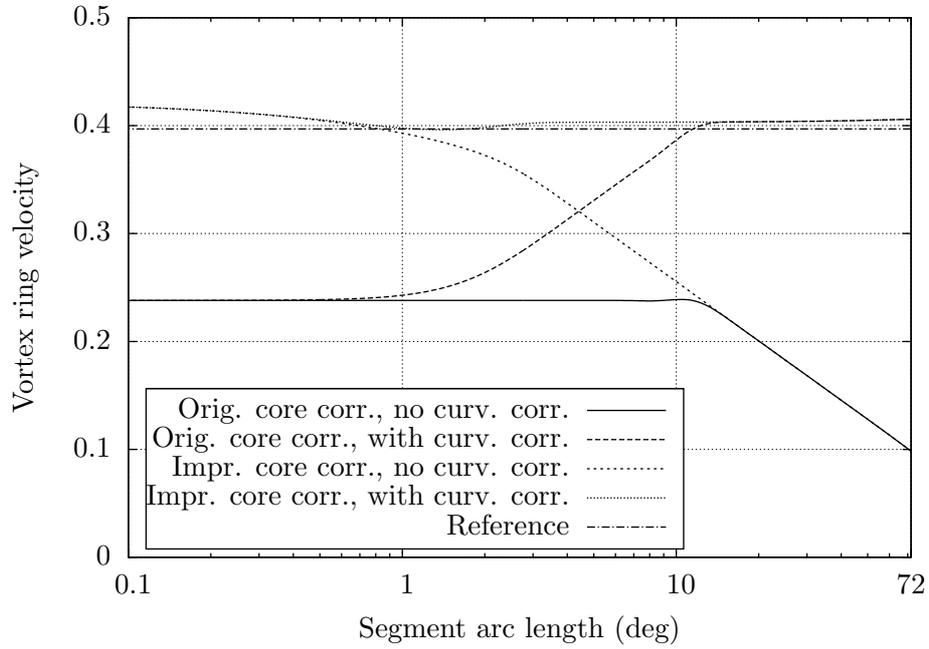}}\hspace{1cm}
    \subfloat[][]{\label{subfig:segm_conv_ring_point_relerr_gaussian}\includegraphics{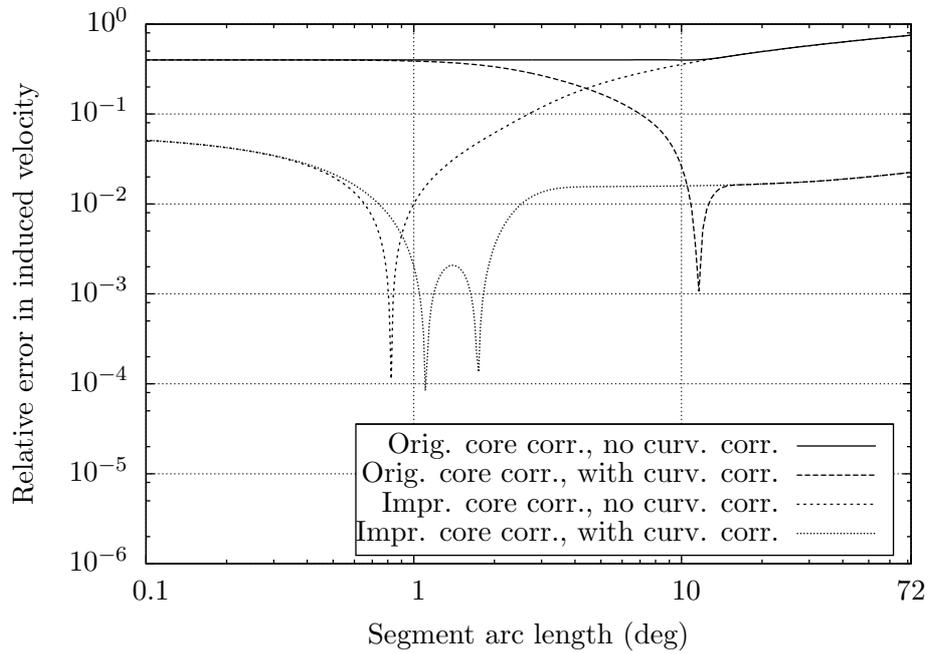}}
    \caption{Induced velocity computed with the segmentation method for a point located on a viscous vortex ring with Gaussian smoothing ($\sigma/R = 0.03$):~\protect\subref{subfig:segm_conv_ring_point_absval_gaussian} absolute values and~\protect\subref{subfig:segm_conv_ring_point_relerr_gaussian} relative errors.}\label{fig:segm_conv_ring_point_gaussian}
\end{figure}

    Results for the point located on the vortex ring are presented in Fig.~\ref{fig:segm_conv_ring_point_gaussian} for a three percent relative core size, again for the normalised Gaussian smoothing, both on an absolute (Fig.~\ref{subfig:segm_conv_ring_point_absval_gaussian}) and relative scale (Fig.~\ref{subfig:segm_conv_ring_point_relerr_gaussian}). Second-order convergence observed as in Fig.~\ref{fig:segm_conv_outside_point_relerr_gaussian} is nowhere to be seen now. The horizontal line at $U_{g3}^{NG}\approx 0.4$ m/s is the reference velocity in this case (Eq.~\ref{eq:vr_numerical_lamb_oseen}). For the four different cases, very different behaviour is observed.
    
    The original core correction model without curvature corrections significantly underpredicts the velocity at coarse discretisations since it neglects the contribution due to the directly adjacent segments. As the discretisation is refined, the value of the computed result increases until at $\Delta\psi\approx 10^{\circ}$ the predicted value levels off and converges to approximately 0.24 m/s. As explained before, the original core correction overcorrects induced velocity values which results in an underprediction of the total induced velocity.
    
    When this method is augmented with a curvature correction, results at coarse discretisations are significantly improved. The correct value is somewhat overpredicted due to discretisation errors. As the discretisation is refined, the magnitude of the curvature correction is reduced and the errors associated with original core correction start to influence the total value. The velocity at very fine discretisations converges to the same value as the previous method as then they are essentially the same. At the discretisation resolution where the computed velocity value crosses the reference value, a small region with a high accuracy results. In the plot with relative errors, this shows as a narrow region where the relative error has a local minimum.
    
    For the improved core correction model without curvature correction, results at coarse discretisations are exactly the same as for the original core correction. As the discretisation is refined, the predicted value keeps increasing until at approximately 0.8 degrees, the correct value is crossed. Further refining the discretisation results in a further increase of the predicted value, to 0.42 m/s at $\Delta\psi=0.1^{\circ}$. From this figure, it is clear that this correction is indeed (just) a correction and not correct (see discussion later). However, the error is significantly reduced with respect to the original core correction.
    
    Combining the improved correction with the curvature correction gives the best overall results. At coarse discretisations, the curvature correction ensures that the predicted value is fairly accurate, while the improved correction prevents the predicted values from decreasing too much. The curve of the predicted values crosses the reference value twice between discretisations of one and two degrees. At finer discretisations, the contribution due to local curvature becomes negligible and the values converge to the values as predicted with the previous method.
    
\begin{figure}
    \centering
    \subfloat[][]{\label{subfig:segm_conv_ring_point_absval_rankine}\includegraphics{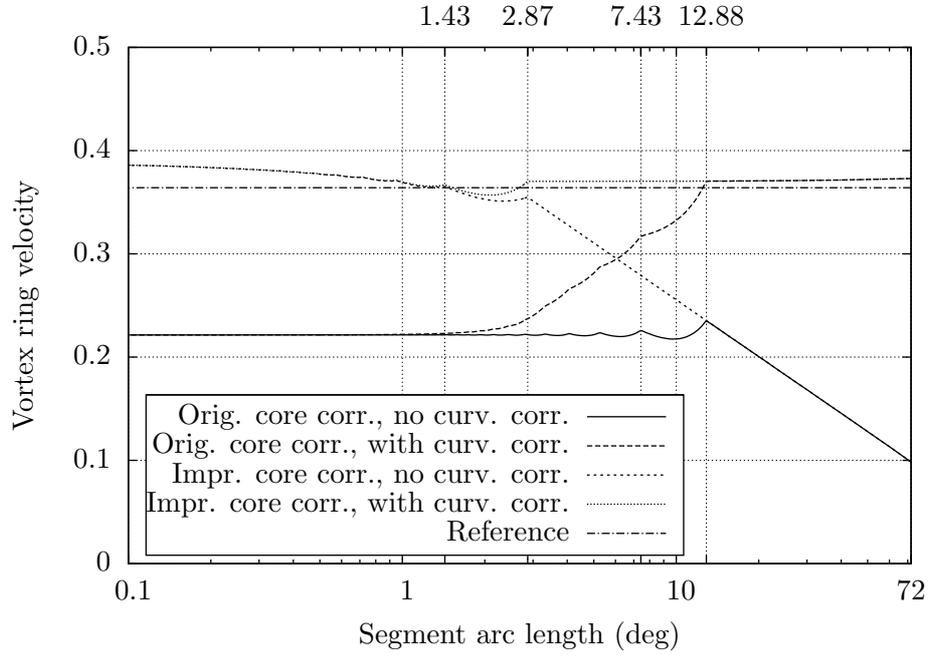}}\hspace{1cm}
    \subfloat[][]{\label{subfig:segm_conv_ring_point_relerr_rankine}\includegraphics{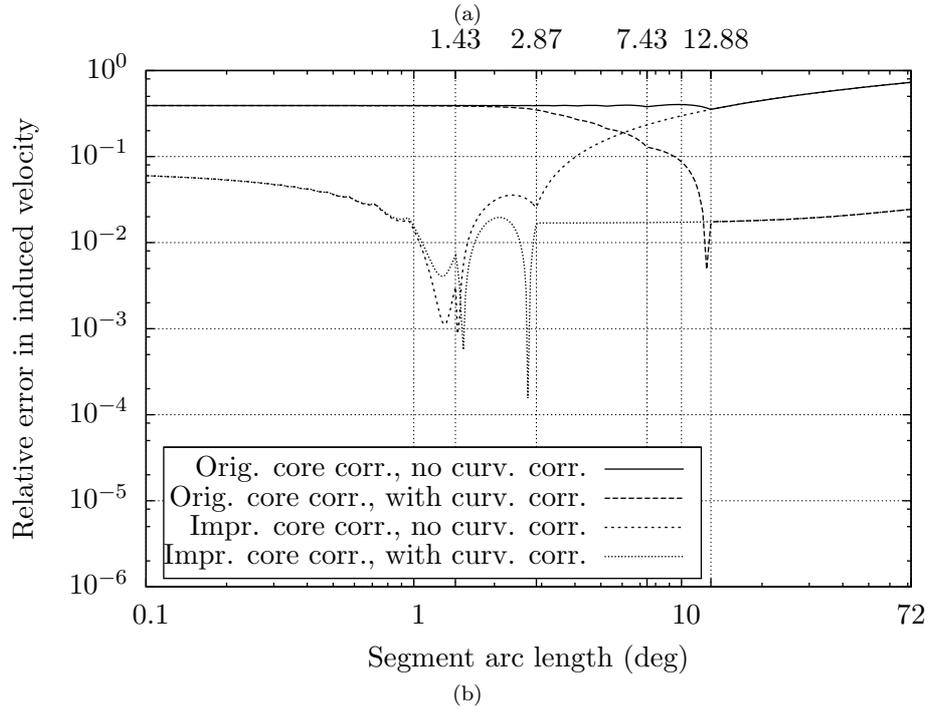}}
    \caption{Induced velocity computed with the segmentation method for a point located on a viscous vortex ring with Rankine smoothing ($\sigma/R = 0.05$):~\protect\subref{subfig:segm_conv_ring_point_absval_rankine} absolute values and~\protect\subref{subfig:segm_conv_ring_point_relerr_rankine} relative errors. Positions of the labels at the upper side are computed with Eqs.~\protect\ref{eq:original_core_correction_kink_prediction} and~\protect\ref{eq:improved_core_correction_kink_prediction}.}\label{fig:segm_conv_ring_point_rankine}
\end{figure}

    For a vortex ring with the solid body rotation smoothing and a relative core size of 5 percent, results are shown in Fig.~\ref{fig:segm_conv_ring_point_rankine}. In this case, the smoothing is not $C^1$ continuous which results in nonsmooth covergence characteristics. The positions of the \emph{kinks} in the curves (both for the original and the improved core correction model) can be predicted analytically due to the sharp boundary between corrected and uncorrected regions for this smoothing.
    
    For the original core correction model, the following implicit equation gives the relationship between the relative core size and the arc length of the segments that results in the position of the \emph{kinks},
    \begin{equation}\label{eq:original_core_correction_kink_prediction}
        \cos\frac{\Delta\psi}{2} - \cos \frac{(2n+1)\Delta\psi}{2} = \frac{\sigma}{R},
    \end{equation}
    where the parameter $n$ is the n\textsuperscript{th} adjacent segment to an evaluation point on a filament that has a nonzero contribution at that evaluation point. Eq.~\ref{eq:original_core_correction_kink_prediction} is implicit in $\Delta\psi$ and must be solved iteratively. A good initial guess for $\Delta\psi = \frac{2}{2n+1}\arccos\left(1-\frac{\sigma}{R}\right)$.
    As an example, for $n=1$ and $n=2$ (which correspond to the first and second adjacent segment with a nonzero contribution at the evaluation point), the values for $\Delta\psi$ equal 12.88\textdegree and 7.43\textdegree, respectively.
    
    For the improved core correction model, the equation for the relationship between the relative core size and the arc length of the segments that gives the position of the kinks is explicit,
    \begin{equation}\label{eq:improved_core_correction_kink_prediction}
        \Delta\psi = \frac{2}{n}\arcsin\left(\frac{\sigma}{2R}\right).
    \end{equation}
    For the first two points, this gives values for $\Delta\psi$ of 2.87\textdegree and 1.43\textdegree. Tick marks are added in Fig.~\ref{fig:segm_conv_ring_point_rankine} at these points.
\section{Discussion}%
\begin{figure}
    \centering
    \subfloat[][]{\label{subfig:nurbs_facet_conv_ring_point_absval_gauss}\includegraphics{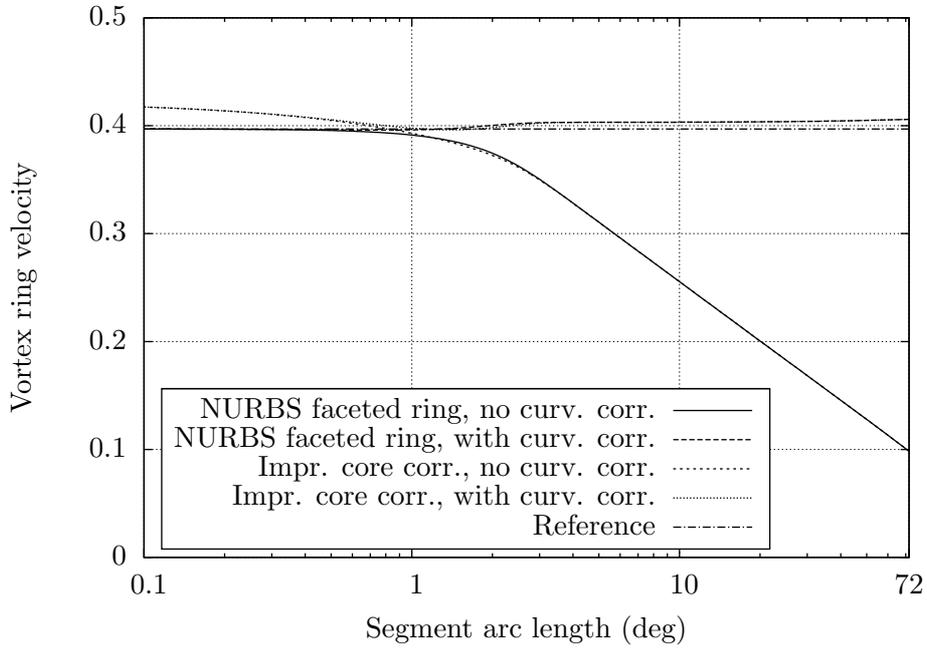}}\hspace{1cm}
    \subfloat[][]{\label{subfig:nurbs_facet_conv_ring_point_relerr_gauss}\includegraphics{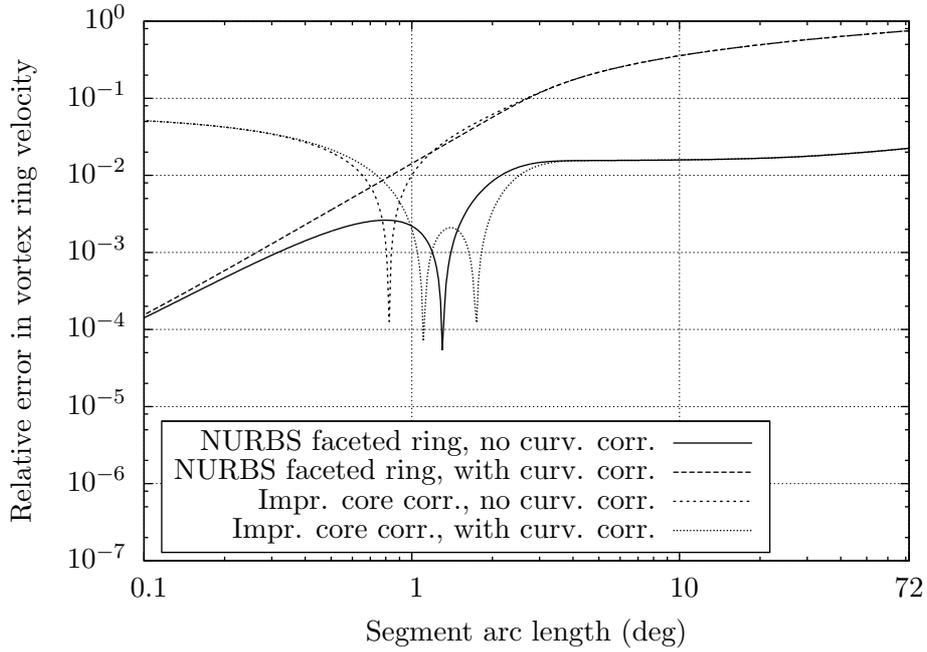}}
    \caption{Induced velocity computed with the segmentation method for a point located on a viscous vortex ring with Gaussian smoothing ($\sigma/R=0.03$) using the improved core correction model (segmented ring) and the numerically computed exact correction (using a NURBS-based regular polygon):~\protect\subref{subfig:nurbs_facet_conv_ring_point_absval_gauss} absolute values and~\protect\subref{subfig:nurbs_facet_conv_ring_point_relerr_gauss} relative errors. }\label{fig:nurbs_facet_conv_ring_point_gauss}
\end{figure}

    The correction proposed in Ref.~\cite{hoydonck_tooren_2011} was shown to give improved convergence of the segmentation method for resolutions to approximately one degree segments, but it is still just a correction. When even finer segment discretisations are desired, the exact correction may be needed. It can be found by correcting the Biot-Savart law with a velocity smoothing before integration instead of correcting it afterwards with the desired swirl velocity profile. As an example, analytic integration of the Biot-Savart law regularised with the Rosenhead-Moore smoothing for a straight line segment,
    \begin{equation}\label{eq:biot_savart_integral_w_rosenhead_moore_smoothing}
        \mb{v}_p^{RM} = -\frac{\Gamma}{4\pi} \int_{C(u)} \frac{\mb{r}}{(|\mb{r}|^2+\sigma^2)^{3/2}} \times \frac{\partial \mb{r}(u)}{\partial u} du
    \end{equation}
    will give the exact correction for Scully's swirl velocity profile. This should then be repeated for each smoothing one wishes to use.
    Using the NURBS-based filament method, it is easy to show that if one would use the analytic result of Eq.~\ref{eq:biot_savart_integral_w_rosenhead_moore_smoothing} with a segmentation method, second-order convergence can be achieved for points on the vortex ring. An exact NURBS representation of a segmented vortex ring (a regular polygon) with $n$ segments is created and the induced velocity is computed at a point on the ring using the regularised Biot-Savart law. Results are shown in Fig.~\ref{fig:nurbs_facet_conv_ring_point_gauss} for the normalised Gaussian smoothing with a three percent core size both on an absolute and relative scale, for both the case without curvature correction and the case including curvature correction. Results for the segmentation method using the improved core correction similar to the results shown before (Fig.~\ref{fig:segm_conv_ring_point_gaussian}) are included for easy comparison. Inspection of Fig.~\ref{subfig:nurbs_facet_conv_ring_point_relerr_gauss} shows that for the case without curvature correction at discretisations finer than $\Delta\psi \approx 3\textrm{\textdegree}$, a second-order trend is observed. 
    The addition of a curvature correction improves the convergence of segmentation methods at coarser discretisations. Results computed using both methods converge to the same values for fine discretisations.
    
    Derivation of the correct, smoothed version of the velocity induced by a straight line segment will only improve the accuracy of results at discretisations finer than what is currently used in engineering wake models, so it may not be worth the effort. Furthermore, the derivation may become more difficult when the vortex strength or the core size is allowed to change along the filament.
    
    Bhagwat and Leishman (Ref.~\cite{bhagwat_leishman_2001_2}) have computed induced velocity values at points near and on a potential and viscous vortex ring for a very fine discretisation and used that as a reference value to compute convergence characteristics at coarser discretisations. The conclusion of Bhagwat and Leishman is that:
    
    \begin{quotation}
        ``Therefore, the discrete straight-line approximation for the induced velocity of either a potential or viscous vortex ring is formally second-order accurate.''
    \end{quotation}
    However, one of the assumptions used to arrive at this conclusion, that the result at the finest discretisation is the correct one and that it can be used as a reference value to compute errors at coarser discretisations, is not true. As shown here, the straight-line approximation method can only attain second-order convergence for points on the ring provided that formulas are corrected before integration, and not afterwards.
\section{Conclusions}%
    The effect of two modeling assumptions used in straight-line segmentation methods have been examined rigorously. The first one is related to the assumption that contributions due to local curvature are sufficiently small so that they may be neglected completely. The second assumption is related to the method used to regularise the analytical solution for the induced velocity due to a single, straight vortex line segment.
    
    A high-order vortex filament method is presented to investigate these modeling assumptions. The method uses Non-Uniform, Rational B-Spline (NURBS) curves as a basis for filament geometry representation. The Biot-Savart law is regularised with a velocity smoothing function and induced velocity values at evaluation points are computed using an adaptive quadrature method. The relationship between three-dimensional velocity smoothing functions and the associated two-dimensional swirl velocity profiles was shown. The three-dimensional smoothings related to various swirl velocity profiles were given except for the swirl velocity of Vatistas ($n=2$), for which it has not been found yet. The method is validated statically by computing the velocity of a viscous vortex ring. For decreasing relative core sizes, numerical results are shown to converge to the value predicted with approximate analytic expressions that assume a circular core shape. These are different from the (correct) expressions where no assumption is made regarding the core shape, except for the Rosenhead-Moore velocity smoothing.
    
    The NURBS-based vortex filament method is subsequently used to study the effects of two common modelling assumptions used in straight-line segmentation methods, the use of core corrections and curvature corrections. For the former one, the original model and an improved version are tested. This analysis shows that for evaluation points located near tangent lines of quadrature points, the original core correction model predicts induced velocity values that are too low. For points on the vortex ring, this results in a consistent underprediction of the ring velocity by approximately 40\%.
    
    The consequences of neglecting the contribution from the adjacent part of a curved filament to the induced velocity are also studied using the NURBS-based vortex filament method. It is shown that for a range of segment discretisation resolutions as used in engineering rotorcraft applications, discarding these corrections gives errors between 1 and 40 percent, depending on the relative core size, the arc length, the particular smoothing and optionally, the vortex Reynolds number used. When the original core correction is used, these errors are orders of magnitude smaller for the same relative core size and arc length.
    
    Finally, the segmentation method is analysed in detail. Induced velocity is computed at three points in the vicinity of a viscous vortex ring. The first one is located at the centre of the ring, the second one is located on the outside and the last point is located on the ring. Results for the former two points are representative for far-field computations while results for the latter point are representative for self-induced velocity computations. It is concluded that for points in the far-field of the vortex ring, second-order convergence can only be achieved with the improved core correction model. For points on the vortex ring, the best possible results over a wide range of discretisations are achieved by combining the curvature correction with the improved core correction. The former ensures that computed values are close to the reference for coarse discretisations while the latter gives good results at fine discretisations. For the improved core correction at discretisations finer than 5 degrees, the use of curvature corrections gives results that are an order of magnitude more accurate. When the discretisations are finer than one degree, differences are so small that curvature corrections may be neglected. Finally, it is demonstrated that the use of corrections applied to the Biot-Savart law after analytical integration precludes second-order convergence to the correct value. This is only possible when the Biot-Savart law is regularised before integrating it analytically.

\bibliography{biblio}

\begin{thebibliography}{10}

\bibitem{abramowitz_stegun_book_1970}
M.~Abramowitz and I.~A. Stegun, editors.
\newblock {\em {Handbook of Mathematical Functions -- with Formulas, Graphs,
  and Mathematical Tables}}.
\newblock {Dover Publications, Inc.}, {New York}, ninth edition, November 1970.

\bibitem{dlmf_nist}
Anon.
\newblock {Digital Library of Mathematical Functions}.

\bibitem{bagai_phd_1995}
A.~Bagai.
\newblock {\em {Contributions to the Mathematical Modeling of Rotor Flow Fields
  Using a Pseudo-Implicit Free-Wake Analysis}}.
\newblock {Ph.D.} thesis, {University of Maryland}, {Maryland, MD}, April 1995.

\bibitem{bhagwat_leishman_2001_2}
M.~J. Bhagwat and J.~G. Leishman.
\newblock {Stability, Consistency and Convergence of Time-Marching Free-Vortex
  Rotor Wake Algorithms}.
\newblock {\em {Journal of the American Helicopter Society}}, 46(1):59--71,
  January 2001.

\bibitem{bliss_teske_quackenbush_1987}
D.~B. Bliss, M.~E. Teske, and T.~R. Quackenbush.
\newblock {A New Methodology for Free Wake Analysis Using Curved Vortex
  Elements}.
\newblock {NASA CR-3958}, {Ames Research Center}, 1987.

\bibitem{castles_leeuw_1954}
W.~Castles~Jr. and J.~H. {De Leeuw}.
\newblock {The Normal Component of the Induced Velocity in the Vicinity of a
  Lifting Rotor and some Examples of its Application}.
\newblock {NACA Report 1184}, {NACA}, 1954.

\bibitem{cottet_koumoutsakos_book_2000}
G.-H. Cottet and P.~Koumoutsakos.
\newblock {\em {Vortex Methods: Theory and Practice}}.
\newblock {Cambridge University Press}, 2000.

\bibitem{meijer_drees_1949}
J.~Meijer Drees.
\newblock {A Theory of Airflow through Rotors and its Application to some
  Helicopter Problems}.
\newblock {\em {The Journal of The Helicopter Association of Great Britain}},
  3(2):79--104, July-Aug.-Sep. 1949.

\bibitem{egolf_1988}
T.~A. Egolf.
\newblock {Helicopter Free Wake Prediction of Complex Wake Structures Under
  Blade{-V}ortex Interaction Operating Conditions}.
\newblock In {\em {Proceedings of the 44\textsuperscript{th} Annual Forum of
  the American Helicopter Society}}, pages 819--832, {Washington, DC}, June
  1988.

\bibitem{fraenkel_1970}
L.~E. Fraenkel.
\newblock {On Steady Vortex Rings of Small Cross-Section in an Ideal Fluid}.
\newblock {\em {Proceedings of the Royal Society of London. Series A,
  Mathematical and Physical Sciences}}, 316(1524):29--62, March 1970.

\bibitem{johnson_camrad_1980}
W.~Johnson.
\newblock {A Comprehensive Analytical Model of Rotorcraft Aerodynamics and
  Dynamics, Part I: Analysis Development}.
\newblock Technical Report NASA-TM-81182, NASA, 1980.

\bibitem{katz_plotkin_book_2001}
J.~Katz and A.~Plotkin.
\newblock {\em {Low Speed Aerodynamics}}.
\newblock {Cambridge University Press}, {Cambridge, UK},
  {2\textsuperscript{nd}} edition, 2001.

\bibitem{leishman_book_2006}
J.~G. Leishman.
\newblock {\em {Principles of Helicopter Aerodynamics}}.
\newblock Cambridge Aerospace Series. {Cambridge University Press},
  {Cambridge}, {2\textsuperscript{nd}} edition, 2006.

\bibitem{leonard_1975}
A.~Leonard.
\newblock {Numerical simulation of interacting three-dimensional vortex
  filaments}.
\newblock In R.~Richtmyer, editor, {\em {Proceedings of the Fourth
  International Conference on Numerical Methods in Fluid Dynamics}}, volume~35
  of {\em {Lecture Notes in Physics}}, pages 245--250. {Springer},
  Berlin/Heidelberg, 1975.

\bibitem{leonard_1985}
A.~Leonard.
\newblock {Computing Three-Dimensional Incompressible Flows with Vortex
  Elements}.
\newblock In {\em {Ann. Rev. Fluid Mech.}}, volume~17, pages 523 -- 559. 1985.

\bibitem{piegl_tiller_1989}
L.~Piegl and W.~Tiller.
\newblock {A Menagerie of Rational B-Spline Circles}.
\newblock {\em {IEEE Computer Graphics and Applications}}, 9(5):48--56,
  September 1989.

\bibitem{piegl_tiller_book_1997}
L.~Piegl and W.~Tiller.
\newblock {\em {The NURBS Book}}.
\newblock {Springer-Verlag}, Berlin, {2\textsuperscript{nd}} edition, 1997.

\bibitem{ramasamy_leishman_2007}
M.~Ramasamy and J.~G. Leishman.
\newblock {A Reynolds Number-Based Blade Tip Vortex Model}.
\newblock {\em {Journal of the American Helicopter Society}}, 52(3):214--223,
  July 2007.

\bibitem{saberi_maisel_1987}
H.~A. Saberi and M.~D. Maisel.
\newblock {A Free-Wake Rotor Analysis Including Ground Effect}.
\newblock In {\em {Proceedings of the 43\textsuperscript{rd} Annual Forum of
  the American Helicopter Society}}, pages 879--889, {St Louis, MO}, 1987.

\bibitem{sadler_1972_1}
S.~G. Sadler.
\newblock {Main Rotor Free Wake Geometry Effects on Blade Air Loads and
  Response for Helicopters in Steady Maneuvers, Volume I -- Theoretical
  Formulation and Analysis of Results}.
\newblock {NASA CR-2110}, {Langley Research Center}, {Langley, VA}, September
  1972.

\bibitem{saffman_1970}
P.~G. Saffman.
\newblock {The Velocity of Viscous Vortex Rings}.
\newblock {\em {Studies in Applied Mathematics}}, 49(4):371--380, December
  1970.

\bibitem{saffman_book_1992}
P.~G. Saffman.
\newblock {\em {Vortex Dynamics}}.
\newblock {Cambridge University Press}, 1992.

\bibitem{scully_1975}
M.~P. Scully.
\newblock {Computation of Helicopter Rotor Wake Geometry and Its Influence on
  Rotor Harmonic Airloads}.
\newblock Technical Report ASRL TR 178-1, {Massachusetts Institute of
  Technology}, {Massachusetts, CA}, March 1975.

\bibitem{shampine_2008}
L.~F. Shampine.
\newblock {Vectorized Adaptive Quadrature in MATLAB}.
\newblock {\em {Journal of Computational and Applied Mathematics}},
  211(2):131--140, February 2008.

\bibitem{hoydonck_gerritsma_tooren_2011}
W.~R.~M. {Van Hoydonck}, M.~I. Gerritsma, and M.~J.~L. {van Tooren}.
\newblock {Vortex Filament Simulation Using NURBS Primitives}.
\newblock In {\em {Proceedings of the 41\textsuperscript{st} AIAA Fluid
  Dynamics Conference and Exhibit}}, number {AIAA-2011-3898}, {Hawaii, HI},
  June 2011.

\bibitem{hoydonck_tooren_2011}
W.~R.~M. {Van Hoydonck} and M.~J.~L. {van Tooren}.
\newblock {Validity of Viscous Core Correction Models for Self-Induced Velocity
  Calculations}.
\newblock {\em {Journal of the American Helicopter Society}}.
\newblock Accepted for publication in the Journal of the American Helicopter
  Society: July 28, 2011.

\bibitem{vatistas_kozel_mih_1991}
G.~H. Vatistas, V.~Kozel, and W.~C. Mih.
\newblock {A Simpler Model for Concentrated Vortices}.
\newblock {\em {Experiments in Fluids}}, 11(1):73--76, April 1991.

\bibitem{winckelmans_phd_1989}
G.~S. Winckelmans.
\newblock {\em {Topics in Vortex Methods for the Computation of Three- and
  Two-Dimensional Incompressible Unsteady Flows}}.
\newblock {Ph.D.} thesis, {California Institute of Technology}, {Pasadena, CA},
  February 1989.

\bibitem{winckelmans_leonard_1993}
G.~S. Winckelmans and A.~Leonard.
\newblock {Contributions to Vortex Particle Methods for the Computation of
  Three-Dimensional Incompressible Unsteady Flows}.
\newblock {\em {Journal of Computational Physics}}, 109(2):247--273, December
  1993.

\end{thebibliography}

\clearpage

\begin{sidewaystable}
    \centering
    \begin{threeparttable}
    \caption{Three- and two-dimensional velocity smoothings and swirl velocity profiles relevant to rotorcraft applications.\label{tab:velocity_smoothings}}
    \begin{tabular}{lccc}
        \toprule
        Smoothing model     & $g_3(\rho)$ & $g_2(\rho)$ & $v_{\theta}(\rho)$ \\
        \midrule
        Rosenhead-Moore\tnote{1} & $\frac{\rho^3}{4\pi(\rho^2+1)^\frac{3}{2}}$ & $\frac{1}{2\pi}\frac{\rho^2}{\rho^2+1}$ & $\frac{\rho}{\rho^2+1}$ \\[2mm]
        Solid Body Rotation, $\rho<1$ & $\frac{\arcsin(\rho) - \rho\sqrt{1-\rho^2}}{2\pi^2}$ & $\frac{\rho^2}{2\pi}$ & $\rho$ \\[2mm]
        Solid Body Rotation, $\rho\ge1$ & $\frac{1}{4\pi}$ & $\frac{1}{2\pi}$ & $\frac{1}{\rho}$ \\[2mm]
        Parametric Gaussian\tnote{2} & $\frac{\textrm{erf}(\rho\sqrt{a}) - 2\rho\sqrt{\frac{a}{\pi}} \exp(-a \rho^2) }{4\pi}$ & $\frac{1 - \exp(-a\rho^2)}{2\pi}$ & $\frac{1 - \exp(-a\rho^2) }{\rho}$ \\[2mm]
        Vatistas (n=2)\tnote{3} & $-$ & $\frac{\rho^2}{2\pi\sqrt{\rho^4+1}}$ & $\frac{\rho}{\sqrt{\rho^4+1}}$ \\[2mm]
        Ramasamy-Leishman   & $\frac{\sum_{n=1}^3 a_n ( \textrm{erf}(\rho\sqrt{b_n}) -2\rho\sqrt{\frac{b_n}{\pi}} \exp(-b_n\rho^2))}{4\pi}$ & $\frac{1-\sum_{n=1}^3 a_n \exp(-b_n \rho^2)}{2\pi}$ & $\frac{1-\sum_{n=1}^{3} a_n \exp(-b_n \rho^2)}{\rho}$\\
        \bottomrule
    \end{tabular}
    \begin{tablenotes}
        \footnotesize
        \item[1] Equivalent of Scully swirl velociy profile.
        \item[2] Equivalent of Lamb-Oseen swirl velociy profile when $a\approx1.2564312$.
        \item[3] Three-dimensional smoothing not known.
      \end{tablenotes}
    \end{threeparttable}
\end{sidewaystable}
\end{document}